\documentclass[aps,11pt,english,tightenlinesletterpaper,nofootinbib,notitlepage]{revtex4-1}

\usepackage{graphicx} 	
\usepackage[justification=justified, singlelinecheck=off, compatibility=false]{caption}
\usepackage{amsmath}	
\usepackage{bm}		
\usepackage{amssymb}	
\usepackage{color} 		
\usepackage{times} 		
\usepackage{tensor} 	
\usepackage{float}
\usepackage{slashed}
\usepackage{hyperref} 
\hypersetup{
    colorlinks=true,       
    linkcolor=red,          
    citecolor=blue,        
    filecolor=magenta,      
    urlcolor=blue           
}


\numberwithin{equation}{section}

\usepackage{myterms}

\newcommand{\AB}{\text{\sc ab}}
\newcommand{\SM}{\text{\sc sm}}
\newcommand{\HS}{\text{\sc hs}}

\newcommand{\qsp}{{\rm qsp}}

\newcommand{\gX}{g_{\text{\sc x}}}

\newcommand{\kup}{|{\bf k}|}
\newcommand{\qup}{|{\bf q}|}

\newcommand{\Ical}{\mathcal{I}}

\newcommand{\Xbb}{\mathbb{X}}

\newcommand{\gpsistr}{g^{\psi \psi}_{\rm str}}
\newcommand{\gHstr}{g^{{\text{\sc h}}}_{\rm str}}
\newcommand{\gHHstr}{g^{\text{\sc hh}}_{\rm str}}
\newcommand{\gZstr}{g^{{\text{\sc z}}}_{\rm str}}
\newcommand{\gZZstr}{g^{{\text{\sc zz}}}_{\rm str}}

\newcommand{\SHint}{S^{{\text{\sc h}}}_{\rm int}}
\newcommand{\SHHint}{S^{\text{\sc hh}}_{\rm int}}
\newcommand{\SZint}{S^{{\text{\sc z}}}_{\rm int}}
\newcommand{\SZZint}{S^{{\text{\sc zz}}}_{\rm int}}

\newcommand{\Scusp}{\mathcal{S}^{\rm (cusp)}}
\newcommand{\Skink}{\mathcal{S}^{\rm (kink)}}
\newcommand{\Skk}{\mathcal{S}^{\rm (k-k)}}
\newcommand{\Tcusp}{\mathcal{T}^{\rm (cusp)}}
\newcommand{\Tkink}{\mathcal{T}^{\rm (kink)}}
\newcommand{\Tkk}{\mathcal{T}^{\rm (k-k)}}

\newcommand{\GHcusp}{\Gamma_{\text{\sc h}}^{\rm (cusp)}}
\newcommand{\GHkink}{\Gamma_{\text{\sc h}}^{\rm (kink)}}
\newcommand{\GHkk}{\Gamma_{\text{\sc h}}^{\rm (k-k)}}
\newcommand{\GHHcusp}{\Gamma_{\text{\sc hh}}^{\rm (cusp)}}
\newcommand{\GHHkink}{\Gamma_{\text{\sc hh}}^{\rm (kink)}}
\newcommand{\GHHkk}{\Gamma_{\text{\sc hh}}^{\rm (k-k)}}
\newcommand{\GZcusp}{\Gamma_{\text{\sc z}}^{\rm (cusp)}}
\newcommand{\GZkink}{\Gamma_{\text{\sc z}}^{\rm (kink)}}
\newcommand{\GZkk}{\Gamma_{\text{\sc z}}^{\rm (k-k)}}
\newcommand{\GZZcusp}{\Gamma_{\text{\sc zz}}^{\rm (cusp)}}
\newcommand{\GZZkink}{\Gamma_{\text{\sc zz}}^{\rm (kink)}}
\newcommand{\GZZkk}{\Gamma_{\text{\sc zz}}^{\rm (k-k)}}
\newcommand{\GABcusp}{\Gamma_{\text{\sc ab}}^{\rm (cusp)}}
\newcommand{\GABkink}{\Gamma_{\text{\sc ab}}^{\rm (kink)}}
\newcommand{\GABkk}{\Gamma_{\text{\sc ab}}^{\rm (k-k)}}

\newcommand{\Com}{\ \ , \quad}
\newcommand{\kap}{\psi}

\DeclareRobustCommand{\SkipTocEntry}[5]{}

\begin{document}

\title{Cosmic Strings in Hidden Sectors:  
1.  Radiation of Standard Model Particles
}

\date{\today}

\author{Andrew J. Long}
\email{andrewjlong@asu.edu}
\author{Jeffrey M. Hyde}
\email{jmhyde@asu.edu}
\author{Tanmay Vachaspati}
\email{tvachasp@asu.edu}
\affiliation{Physics Department, Arizona State University, Tempe, Arizona 85287, USA.}

\begin{abstract}

In hidden sector models with an extra $\U{1}$ gauge group, new fields can interact with the Standard Model only through gauge kinetic mixing and the Higgs portal. After the $\U{1}$ is spontaneously broken, these interactions
couple the resultant cosmic strings to Standard Model particles.  
We calculate the spectrum of radiation emitted by these ``dark strings'' in the form of Higgs bosons, Z bosons, and Standard Model fermions assuming that string tension is above the TeV scale.  
We also calculate the scattering cross sections of Standard Model fermions on dark strings due to the Aharonov-Bohm interaction.  
These radiation and scattering calculations will be applied in a subsequent paper to study the cosmological evolution and observational signatures of dark strings.  

\end{abstract}

\maketitle

\setlength{\parindent}{40pt}
\setlength{\parskip}{-0.20cm}
\begingroup
\hypersetup{linkcolor=black}
\tableofcontents
\endgroup
\setlength{\parskip}{0.2cm}

\section{Introduction}\label{sec:Introduction}

In the Standard Model of particle physics, all of the matter fields are charged under the gauge group of the theory, and consequently all of the particles participate in the gauge interactions.  
It is natural to ask whether there can be new particles that do not participate in any of the Standard Model gauge interactions, and whose fields are singlets under the Standard Model gauge group.  
Such fields would be sequestered in a ``hidden sector'' where they participate in their own gauge interactions under which the SM fields are singlets.  
Despite their minimal nature, hidden sector models admit a rich phenomenology; they have been well-studied in the context of collider physics \cite{Leike:1998wr, Langacker:2008yv, Djouadi:2012zc, Alves:2013tqa, No:2013wsa} as well as dark matter \cite{ArkaniHamed:2008qp, Cassel:2009pu, Chun:2010ve, Chu:2011be, Baek:2014jga, Basak:2014sza}.  
In Refs.~\cite{Vachaspati:2009jx,Hyde:2013fia} we have pointed out that these models may also contain cosmic string solutions, called ``dark strings'', that have novel interactions with Standard Model fields.
The aim of the present paper is to derive the radiative and scattering properties of these strings.
In a subsequent paper we will use these properties to study potential astrophysical and cosmological signatures of dark strings.

The Lagrangian for the hidden sector model under consideration is of the form
\begin{equation}
	\Lcal = \Lcal_{\SM} + \Lcal_{\HS} + \Lcal_{\rm int} \per
\end{equation}
The first term, $\Lcal_{\SM}$, is the Standard Model (SM) Lagrangian; the second term, 
\begin{equation}\label{eq:LHS}
	\Lcal_{\HS} = | D_\mu S|^2 - \frac{1}{4} {\hat X}_{\mu\nu}^2 - \kappa ( S^*S - \sigma^2)^2 \com
\end{equation}
is the hidden sector (HS) Lagrangian with $S$ a complex scalar field charged under a $\U{1}_X$ gauge group that has ${\hat X}_{\mu}$ as its gauge potential, $D_\mu = \partial_\mu - i \gX {\hat X}_\mu$; and the third term, 
\begin{equation}\label{eq:Lint}
	\Lcal_{\rm int} = -\alpha (\Phi^{\dagger} \Phi -\eta^2)( S^{\ast} S-\sigma^2) - \frac{\sin\epsilon}{2} {\hat X}_{\mu\nu}Y^{\mu\nu} \com
\end{equation}
is the interaction Lagrangian with $\Phi$ the Higgs doublet and $Y_{\mu}$ the hypercharge gauge field.  
The mass scale of the hidden sector fields is set by the parameter $\sigma$, and $\eta = 174 \GeV$ is the vacuum expectation value (VEV) of the Higgs field.  
The two terms in $\Lcal_{\rm int}$ are called the Higgs portal (HP) term \cite{Patt:2006fw} and the gauge-kinetic mixing (GKM) term \cite{Holdom:1985ag, Foot:1991kb}, respectively.  
For $\sigma \lesssim {\rm TeV}$, the HP and GKM couplings are well-constrained, $|\alpha|, |\sin \epsilon| \ll 1$ \cite{Jaeckel:2013ija,Belanger:2013kya}, but if $\sigma$ is above the TeV scale, making HS particles inaccessible at laboratory energies, the hidden sector model is (as yet) unconstrained.  
In principle the hidden sector can be extended to include additional fields and interactions; we retain only the minimal degrees of freedom necessary to study radiation of SM particles from the cosmic string.

The VEV $\langle S \rangle = \sigma$ spontaneously breaks the $\U{1}_X$ completely.  
Consequently the model admits topological (cosmic) string solutions \cite{VilenkinShellard:1994}.  
The string tension is set by the symmetry breaking mass scale $\mu \approx \sigma^2$, and we will use $M \equiv \sqrt{\mu} \approx \sigma$ through the text.  
In \rref{Hyde:2013fia} we studied these ``dark string'' solutions, which were found to contain a non-trivial structure in the dark sector fields, $S$ and $\hat{X}_\mu$, as well as in the SM fields, $\Phi$ and $Y_\mu$.  (See also \cite{Peter:1992} for the case when $\sin \epsilon = 0$.)  
In the decoupling limit, $\sigma \gg \eta$, the dark fields form a thin core of thickness on the order of $\sigma^{-1}$, and the SM fields form a wide dressing with thickness $\eta^{-1}$.  
The dressing arises because the string core sources the SM Higgs and Z boson fields, $\phi_H$ and $Z_{\mu}$.  
In the limit $\sigma \gg \eta$ we can integrate out the heavy HS fields leaving only the zero thickness string core.  
In \rref{Hyde:2013fia} we found the effective interaction of the string core with the light SM fields to be 
\begin{equation}\label{eq:Sint}
S_{\rm int}^{(1)} = \gHstr \, \eta   \, \int d^2\sigma \sqrt{-\gamma} \ \phi_H({\mathbb X}^\mu) 
                    + \frac{\gZstr}{2} \left( \frac{\eta}{\sigma} \right)^2 \int d\sigma^{\mu\nu} Z_{\mu\nu}({\mathbb X}^{\mu})
\end{equation}
where $\Xbb^{\mu}$ denotes the location of the zero thickness string core, and the rest of the notation is defined in \aref{app:Worldsheet}.  
The coupling constants $\gHstr$ and $\gZstr$ have been derived in \rref{Hyde:2013fia} in
terms of $\alpha$, $\sin \epsilon$, and other Lagrangian coupling constants.  
We shall treat them as free parameters in the present paper.  
Note that the interaction in \eref{eq:Sint} is valid for $\sigma \gg \eta$, when the string core is much thinner than the SM dressing. 
If the core and dressing widths are comparable, the effective interaction formalism breaks down and the full field theory equations must be solved to evaluate string-particle interactions.

The linear interactions given above arise because the Higgs gets a VEV, and the string acts as a source that modifies the VEV.  
In addition, the string also couples to the SM fields through the more generic quadratic interactions.  
Upon integrating out the heavy hidden sector fields, the effective quadratic interactions for the Higgs and Z boson are
\begin{equation}\label{eq:Sintquadratic}
S_{\rm int}^{(2)} = 
	\gHHstr  \, \int d^2\sigma \sqrt{-\gamma} \ \phi_H^2(\Xbb)
	+ \gZZstr \left( \frac{\eta}{\sigma} \right)^4 \int d^2\sigma \, \sqrt{-\gamma} \, Z_{\mu}(\Xbb) Z^{\mu}(\Xbb) \per 
\end{equation}
The quadratic Higgs interaction derives directly from the HP term in \eref{eq:Lint}, and we can estimate $\gHHstr \approx \alpha$ up to order one factors related to integrals of the profile functions.  
The quadratic Z boson interaction results from the mixing of the Z boson with the heavy $\hat{X}^{\mu}$ field.  
The mixing angle goes like $(\sin \epsilon) (\eta / \sigma)^2$ \cite{Hyde:2013fia}, and therefore we obtain the quadratic interaction in \eref{eq:Sintquadratic} with $\gZZstr \approx \sin^2 \epsilon$.  
The W bosons will have a coupling similar to the Z boson coupling in \eref{eq:Sintquadratic}, and our results for the Z bosons carry over to the other weak bosons as well.  
The remaining bosonic SM fields, the gluons and the photons, do not couple to the string worldsheet at leading order \cite{Hyde:2013fia}.

In addition to interactions with $\phi_H$ and $Z_\mu$, the string also couples to the SM fermions
due to an Aharonov-Bohm (AB) interaction \cite{Alford:1988sj}.  
Upon circumnavigating the string on a length scale larger than the width of the SM dressing fields, the fermion wavefunction picks up a phase that is $2 \pi$ times \cite{Hyde:2013fia}
\begin{equation}\label{thetaq}
\theta_q = - \frac{2 \cos \theta_W \, \sin \epsilon}{g_X} q .
\end{equation}
where $q$ is the electromagnetic charge on the fermion, and $\theta_W$ is the weak mixing angle.  
The AB interaction is topological, insensitive to the details of the structure of the string, and in particular, does not assume $\sigma \gg \eta$.

 \begin{figure}
\begin{center}
\includegraphics[width=0.3\textwidth]{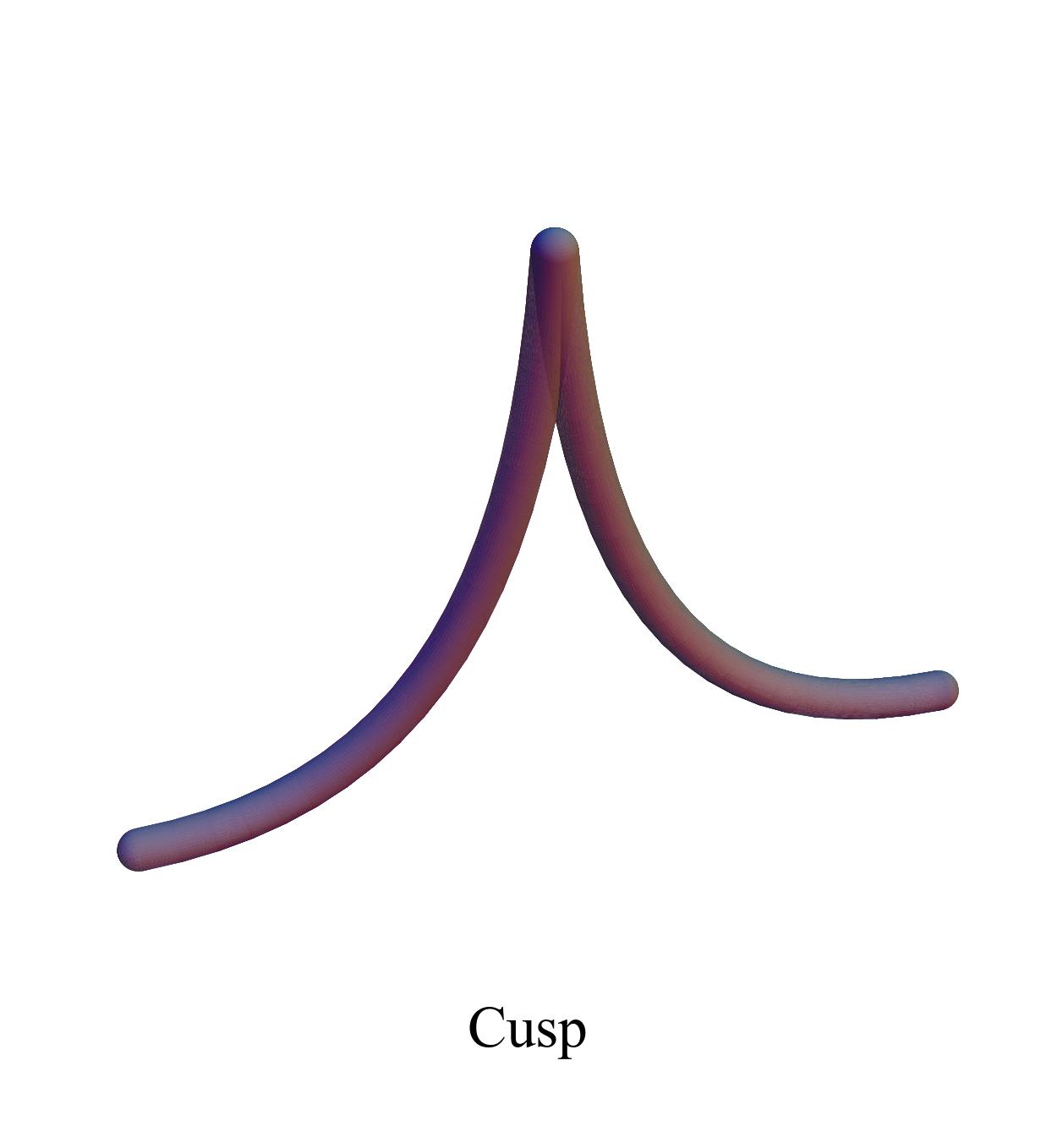} \hfill
\includegraphics[width=0.3\textwidth]{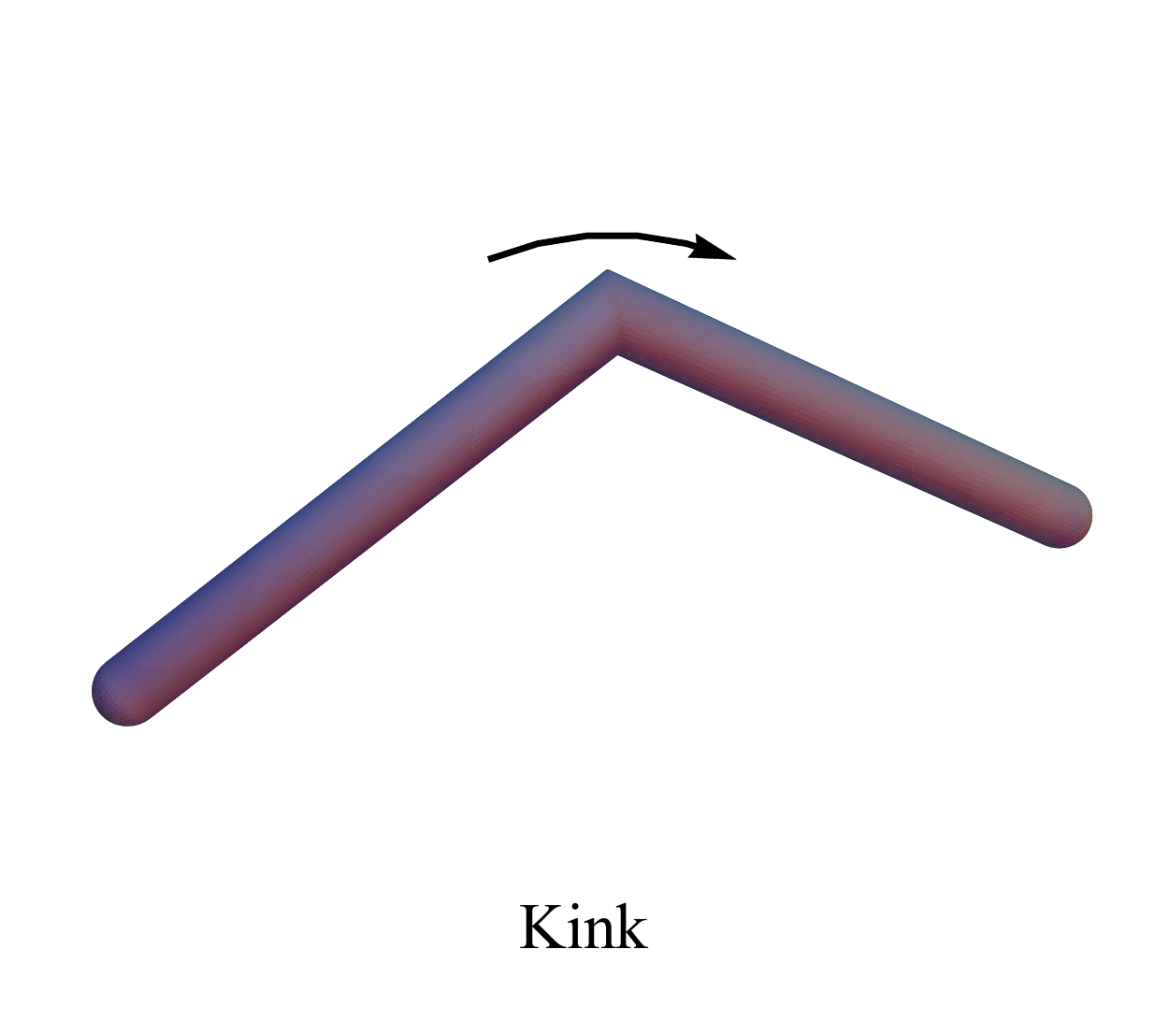} \hfill
\includegraphics[width=0.3\textwidth]{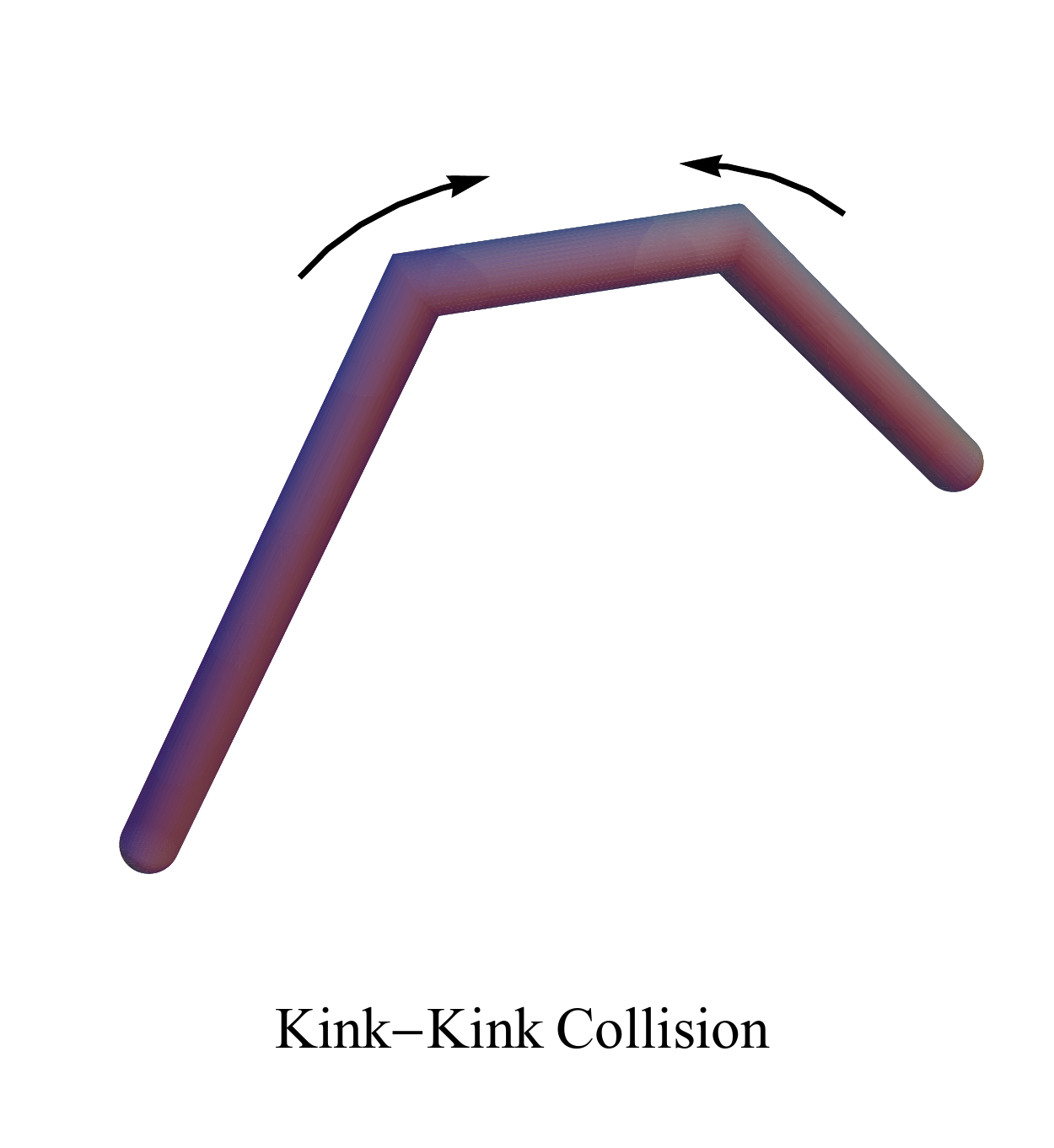} 
\caption{
\label{cuspetc}
Illustrations of the cusp, kink, and kink-kink collision. 
A cusp is a
point on the loop that instantaneously moves at the speed of light; a kink is a discontinuity
in the tangent vector to the string that moves around the loop in one direction; a kink-kink
collision occurs when two oppositely moving kinks collide.}
\end{center}
\end{figure}

By virtue of the interactions in \erefs{eq:Sint}{eq:Sintquadratic}, dynamical dark strings will emit Higgs and Z bosons, and it will emit SM fermions through the AB interaction.  
In the following sections, we calculate the spectrum of radiation in the form of Higgs and Z bosons that is emitted from cusps, kinks, and kink collisions on cosmic string loops (see Fig.~\ref{cuspetc}).  
The scalar boson radiation channels have been derived previously \cite{Srednicki:1986xg, Damour:1996pv, Vachaspati:2009kq, Sabancilar:2009sq}.  
We refine these calculations by carefully estimating all dimensionless coefficients, and in some cases also correcting errors.  
Most importantly, we find that the calculation of \rref{Srednicki:1986xg} underestimates the scalar radiation by a factor of $\sqrt{ML} \gg 1$, which arises because the radiation from the cusp is highly boosted.  
The vector boson channels have not been worked out previously, and we present them here for the first time.  
We also estimate radiation from the Aharonov-Bohm interaction by drawing on results from the literature.  
Our results, it should be emphasized, are not unique to the dark string model; instead, the spectra derived here apply to any model with effective interactions of the form in \erefs{eq:Sint}{eq:Sintquadratic}.  

Particle radiation is expected to play an important role in the evolution of light cosmic string for which gravitational radiation is suppressed.  
Specifically, we find that Higgs boson emission is the dominant energy loss mechanism for light dark strings.  
The emission of SM particles may also lead to observational signatures of dark strings through astrophysics or cosmology, and we will explore this  possibility in a companion paper \cite{Longetal:2014}.

\section{Radiation of Standard Model Particles}\label{sec:SM_Particle_Prod}

The interactions in \erefs{eq:Sint}{eq:Sintquadratic} allow a dark string to emit Higgs and Z bosons, and SM fermions are radiated by virtue of the non-local Aharanov-Bohm interaction.  
In the subsections below we first present the spectrum of Higgs and Z boson radiation from a general string configuration, and we then specify to the cases of cusps, kinks, and kink-kink collisions as these are expected to the be the three most copious sources of particle radiation.  
We leave the details of these calculations to the Appendices.

\subsection{Higgs Boson Emission via Linear Coupling}\label{sub:HiggsEmission_Linear}

The physical Higgs field, $\phi_H(x)$, couples to the dark string through the effective interaction
\begin{equation}\label{eq:Sint_H}
	\SHint = \gHstr \, \eta   \, \int d^2\sigma \sqrt{-\gamma} \ \phi_H(\Xbb) \per
\end{equation}
Since this term is linear in $\phi_H$ it acts as a classical source term for the Higgs field and leads to radiation from the string.  
Note that the dimensional prefactor, $\eta \approx 174 \GeV$, is the vacuum expectation value of the Higgs field;  this interaction would not be present if not for electroweak symmetry breaking.  
In \aref{app:Radiation} we calculate the spectrum of Higgs boson radiation for a string loop.  
Taking $A = \gHstr \eta$ in \eref{eq:Hsingle_dN} we find 
\begin{align}\label{eq:dNH_of_I}
	dN_H &= (\gHstr \eta)^2 \abs{\Ical(k) }^2 \frac{\kup d\omega d\Omega}{2(2\pi)^3} 
\end{align}
where the integral 
\begin{align}
	&  \Ical(k) = \int d^2 \sigma \, \sqrt{-\gamma} \, e^{i k \cdot \Xbb} \, ,
\end{align}
is a functional of the string worldsheet, $\Xbb^{\mu}(\tau,\sigma)$, that describes the motion of the string loop.  
The kinematical variables are defined by $k^{\mu} = \bigl\{ \omega \, , \, {\bf k} \bigr\}$ with $\omega = (m_H^2 + \kup^2)^{1/2}$ 
and $m_H$ the Higgs boson mass.  
In the following subsections we specify $\Xbb^{\mu}$ so as to evaluate the spectrum and total power of Higgs boson emission from cusps, kinks, and kink-kink collisions.  

\addtocontents{toc}{\SkipTocEntry}
\subsubsection{Higgs Emission from a Cusp}\label{sub:HiggsCusp}

A cusp occurs when there is a point on the worldsheet where $\partial_{\sigma} \Xbb = 0$.  
At this point the velocity of the string segment approaches the speed of light, and the radiation is highly boosted.  
In the rest frame of the loop, the momentum of the emitted radiation cannot exceed the inverse string thickness, {\it i.e.} $\kup < M$ where $M = \sqrt{\mu}$, else the point-like interaction in \eref{eq:Sint_H} is inapplicable, and the radiation is suppressed.  
However, due to the large boost factor, $\gamma_{\rm boost} \sim \sqrt{ML}$, the radiation does not cut off until $\kup \approx M \sqrt{ML}$ \aref{app:Boost}.  

Inserting the scalar integral from \eref{eq:I_scalar_cusp} into the spectrum in \eref{eq:dNH_of_I} we find 
\begin{align}\label{eq:dNH_cusp_1}
	& dN_H^{\rm (cusp)} = 
	\frac{(\gHstr \eta)^2}{2 (2\pi)^3} \Scusp
	\frac{L^{4/3}}{\kup^{5/3}} \, d\omega \, d\Omega \ , \nn
	& \qquad
	\kap \, m_H \sqrt{m_H L} < \kup < M \sqrt{ML}
	\, , \quad
	\theta < \Theta \, (\kup L)^{-1/3} \, {\rm (cone)} \per
\end{align}
where $\kap \approx 0.01$ (see \eref{eq:kmin_bound}), $\Theta \approx 0.1$ (below \eref{eq:sp_phase}), and $0.2 \lesssim \Scusp \lesssim 10$ (see below \eref{eq:I_scalar_cusp}).
As explained above, the spectrum is cutoff in the UV by the (boosted) string thickness, and it cuts off in the IR due to a destructive interference that is manifest in the breakdown of the saddle point approximation. 
Since typically $m_H L \gg 1$, the radiation is ultra-relativistic
and we can approximate $\kup \approx \omega$ and $d\kup \approx d\omega$.  

The radiation is emitted into a cone that has an opening angle $\Theta (\kup L)^{-1/3}$.  
Integrating over the solid angle, we find the spectrum to be 
\begin{align}\label{eq:dNH_cusp}
	dN_{\rm H}^{\rm (cusp)}
	\approx 
	\frac{(\gHstr \eta)^2}{4 (2\pi)^2} \Theta^2 \Scusp L^{2/3} \frac{d\kup}{\kup^{7/3}} 
	\, , \quad 
	\kap \, m_H \sqrt{m_H L} < \kup < M \sqrt{ML} \per
\end{align}
The total energy emitted from a cusp is
\begin{align}\label{eq:EH_cusp}
	E_H^{\rm (cusp)} 
	= \int \omega \, dN_H^{\rm (cusp)} 
	= \frac{3(\gHstr \eta)^2}{4 (2\pi)^2} \kap^{-1/3} \Theta^2 \Scusp \sqrt{ \frac{L}{m_H} } \left( 1 - \kap^{1/3} \sqrt{\frac{m_H}{M}} \right) \per
\end{align}
Since we are interested in heavy strings, $M \gg m_H$, we can neglect the second term in the parenthesis.  
If cusps appear on a loop with frequency $f_c / T$ where $T = L / 2$ is the loop oscillation period, then the average power emitted per oscillation is $P_H = 2 E_H f_{c} / L$, or
\begin{align}\label{eq:PH_cusp}
	P_H^{\rm (cusp)} = \GHcusp \frac{(\gHstr \eta)^2}{\sqrt{m_H L}} 
\end{align}
where $\GHcusp \equiv \frac{3}{2(2\pi)^2} \kap^{-1/3} \Theta^2 f_c \, \Scusp$.  
Assuming $f_c \approx 1$, the dimensionless coefficient takes values in the range $10^{-4} \lesssim\GHcusp \lesssim 10^{-1}$.  
This result agrees with a previous calculation in the literature \cite{Vachaspati:2009kq}.  

\addtocontents{toc}{\SkipTocEntry}
\subsubsection{Higgs Emission from a Kink}\label{sub:HiggsKink}

A kink occurs where there is a discontinuity in the derivative of the string worldsheet $\partial_{\sigma} \Xbb$. 
We obtain the spectrum of Higgs radiation emitted from a single kink over the course of one loop oscillation period by evaluating \eref{eq:dNH_of_I} with \eref{eq:I_scalar_kink}, and we find 
\begin{align}\label{eq:dNH_kink}
	dN_{\rm H}^{\rm (kink)} =  & 
	\frac{(\gHstr\eta)^2}{2 (2\pi)^3} \Skink \frac{L^{2/3}}{\kup^{7/3}} \, 
	d\kup \, d\Omega
	\com \nn & 
	\kap \, m_H \sqrt{m_H L} < \kup < M
	\Com
	\theta < \Theta \, (\kup L)^{-1/3} \, {\rm (band)}
\end{align}
where the dimensionless coefficient is typically in the range $0.05 \lesssim \Skink \lesssim 10$.  
Here the upper bound on $k$ is $M$, rather than $M\sqrt{ML}$ as for the cusp, since the string velocity at the kink is not highly boosted in the loop's rest frame.  
The lower bound on $k$ is the same as in the case of the cusp as it arises from our use of the saddle point approximation in one of the worldsheet integrals $\Ical_\pm$ (see Appendix~\ref{app:Worldsheet_Integrals}).  
Unless the loop is very small, $L < M^2 / m_H^3$, the lower cutoff will exceed the upper cutoff; in this case, there is no Higgs radiation from the kink within our approximations.  
This argument is in contrast with the calculation of \rref{Lunardini:2012ct}, where scalar radiation from the kink was also studied.  

Radiation is emitted into a band that has an angular width $\Theta (\kup L)^{-1/3}$ and angular length $\sim 2\pi$.  
Integrating over the sold angle $\Delta\Omega \approx 2\pi \Theta (\kup L)^{-1/3}$ gives the spectrum 
\begin{align}\label{eq:dNH_kink}
	dN_{\rm H}^{\rm (kink)} =  
	\frac{(\gHstr\eta)^2}{2(2\pi)^2} \, \Theta \, \Skink \, L^{1/3} \, \frac{d\kup}{\kup^{8/3}} 
	\, , \qquad
	\kap \, m_H \sqrt{m_H L} < \kup < M \per
\end{align}
The total energy emitted by the kink into this channel during one loop oscillation is
\begin{align}\label{eq:EH_kink}
	E_H^{\rm (kink)} = \int \omega \, dN_H^{\rm (kink)} 
	= \frac{3 (\gHstr\eta)^2}{4(2\pi)^2} \, \psi^{-2/3} \Theta \, \Skink \, \frac{1}{m_H} \left( 1 - \psi^{2/3} \frac{m L^{1/3}}{M^{2/3}} \right)
\end{align}
Note that the energy is logarithmically sensitive to both the upper and lower cutoffs of the spectrum.  
If the loop carries $N_k$ kinks, then the average power radiated during one loop oscillation period, $T = L /2$, is given by 
\begin{align}\label{eq:PH_kink}
	P_H^{\rm (kink)} = \GHkink \frac{(\gHstr\eta)^2}{m_H L} \left( 1 - \psi^{2/3} \frac{m_H L^{1/3}}{M^{2/3}} \right)
\end{align}
with $\GHkink = \frac{3}{2(2\pi)^2} N_{k} \, \psi^{-2/3} \Theta \, \Skink$. 
Taking $N_k \approx 1$ the dimensionless prefactor is estimated to be $10^{-3} \lesssim \GHkink \lesssim 1$.  
This result disagrees with a previous calculation \cite{Lunardini:2012ct} of Higgs radiation from a 
kink, as explained in \aref{sub:ScalarKink}.  

\addtocontents{toc}{\SkipTocEntry}
\subsubsection{Higgs Emission from a Kink-Kink Collision}\label{sub:HiggsKinkKink}

A kink-kink collision occurs when two kinks momentary overlap at the same point on the string worldsheet.  
We find the spectrum of Higgs radiation at the collision using \eref{eq:dNH_of_I} along with the scalar integral in \eref{eq:I_scalar_kk}:
\begin{align}\label{eq:dNH_kk}
	dN_{\rm H}^{\rm (k-k)} =  \frac{(\gHstr\eta)^2}{2 (2\pi)^3}\frac{\Skk}{\omega^4} \, \kup \, d\omega \, d\Omega 
	\Com
	m_H  < \omega < M 
\end{align}
where $0.05 < \Skk < 200$.  
The bound $\omega > m_H$ subsumes the bound $\omega > L^{-1}$ in \eref{eq:I_scalar_kk} assuming $m_H L \gg 1$.  

The radiation is emitted approximately isotropically (no beaming), and the angular integration gives 
\begin{align}\label{eq:dNH_kk}
	dN_{\rm H}^{\rm (k-k)} =  \frac{(\gHstr\eta)^2}{(2\pi)^2} \Skk \frac{\kup}{\omega^4} \, d\omega 
	\Com
	 m_H < \omega < M \per
\end{align}
The total energy emitted by a kink-kink collision is found to be 
\begin{align}\label{eq:EH_kk}
	E_H^{\rm (k-k)} = \int \omega \, dN_H^{\rm (k-k)} 
	                          = \frac{(\gHstr\eta)^2}{(2\pi)^2} \, \Skk \frac{1}{m_H} .
\end{align}
Defining $N_{kk}$ as the number of kink-kink collisions during one loop oscillation period, $T = L / 2$, we can express the average power radiated by
\begin{align}\label{eq:PH_kk}
	P_H^{\rm (k-k)} & = \GHkk \frac{(\gHstr\eta)^2}{m_H L} 
\end{align}
with $\GHkk \equiv \frac{2}{(2\pi)^{2}} N_{kk} \, \Skk$.  
We can estimate the number of collisions per loop oscillation period as $N_{kk} \approx N_k^2$, where $N_k$ is the number of kinks on the loop.
Estimating $N_{kk} \approx 1$ we obtain a range $10^{-2} < \GHkk < 10$ for the dimensionless prefactor.

\subsection{Higgs Boson Emission via Quadratic Coupling}\label{sub:HiggsPairEmission_Quadratic}

The radial component of the Higgs field also couples to the dark string through the quadratic interaction 
\begin{equation}\label{eq:SHH_int}
	\SHHint = \gHHstr  \, \int d^2\sigma \sqrt{-\gamma} \ \phi_H^2(\Xbb) \per 
\end{equation}
Unlike in the linear type coupling discussed above, this interaction is not proportional to the Higgs field VEV, and it would exist even if the electroweak symmetry were unbroken.  
This quadratic interaction with the string produces two Higgs bosons, and thus the final state contains
two different momenta, $k$ and ${\bar k}$.
The spectrum of radiation is given by \eref{eq:Hquad_dN} with $C = \gHHstr$:  
\begin{align}\label{eq:dNHH_of_I}
	dN_{HH} &= (\gHHstr)^2 \frac{\kup \, d\omega \, d\Omega}{2(2\pi)^3}  \frac{\bar{\kup} \, d\bar{\omega} \, d\bar{\Omega}}{2(2\pi)^3}  \abs{\Ical(k + \bar{k}) }^2
\end{align}
where $k^{\mu} = \bigl\{ \omega \, , \, {\bf k} \bigr\}$ with $\omega = (m_H^2 + \kup^2)^{1/2}$
and $m_H$ the Higgs boson mass.  
The barred quantities are defined similarly.  

\addtocontents{toc}{\SkipTocEntry}
\subsubsection{Higgs-Higgs Emission from a Cusp}\label{sub:HiggsQuadCusp}

Before we can evaluate the spectrum in \eref{eq:dNHH_of_I} we must know the value of the scalar integral $\Ical(k+\bar{k})$ for a cusp configuration.  
In \eref{eq:I_scalar_cusp} we found that this integral evaluates to 
\begin{align}\label{eq:Hquad_Icusp_before}
	\bigl| \mathcal{I}^{\rm (cusp)} (k) \bigr|^2 = \Scusp \frac{L^{4/3}}{\kup^{8/3}} 
	\Com
	\kap \, m_H \sqrt{m_H L} < \kup 
	\Com
	\theta < \Theta (\kup L)^{-1/3} 
\end{align}
when its argument is the approximately null 4-vector momentum $k^2 = m_H^2 \ll \kup^2$.  
If the argument of the integral is a time-like vector, as in \eref{eq:dNHH_of_I}, the derivation still leads to \eref{eq:Hquad_Icusp_before}, but the saddle point approximation gives an additional bound on the angle between ${\bf k}$ and $\bar{\bf k}$.  
In order to justify the saddle point approximation, we were forced to impose the bound in \eref{eq:general_bound}.  
Since the argument of the integral in \eref{eq:dNHH_of_I} is $k + \bar{k}$, we must generalize \eref{eq:general_bound} by replacing $\omega \to \omega + \bar{\omega}$ and $\kup \to | {\bf k} + \bar{\bf k} | = \sqrt{ \kup^2 + \bar{\kup}^2 + 2 \kup \bar{\kup} \cos \theta_{k \bar{k}} }$ where $\theta_{k\bar{k}}$ is the angle between ${\bf k}$ and $\bar{\bf k}$.  
The bound becomes
\begin{align}
	\frac{\Theta}{4\pi} L^{2/3} (\omega + \bar{\omega} - \sqrt{ \kup^2 + \bar{\kup}^2 + 2 \kup \bar{\kup} \cos \theta_{k \bar{k}} } ) < ( \kup^2 + \bar{\kup}^2 + 2 \kup \bar{\kup} \cos \theta_{k \bar{k}} )^{1/6} \per
\end{align}
It is useful to consider two limiting cases.  
If $\theta_{k \bar{k}} = 0$ then the inequality translates into a lower bound on the momentum, 
\begin{align}
	\kap \, m_H \sqrt{m_H L} < \frac{(\kup \bar{\kup})^{3/4}}{\sqrt{\kup + \bar{\kup}}} \com
\end{align}
and we have used $\kap = (\Theta / 8 \pi)^{3/4}$.  
When $\kup \approx \bar{\kup}$ we regain the original bound $\kap \, m \sqrt{mL} < \kup , \bar{\kup}$.  
The inequality also imposes an upper bound on $\theta_{k \bar{k}}$.  
Approximating $\cos \theta_{k \bar{k}} \approx 1 - \theta_{k \bar{k}}^2 / 2$ and using $\omega \approx \kup$ and $\bar{\omega} \approx \bar{\kup}$, we can resolve the inequality as
\begin{align}\label{eq:theta_kk}
	\theta_{k \bar{k}} < \kap^{-2/3} \frac{(\kup + \bar{\kup})^{2/3} L^{-1/3} }{\sqrt{\kup \bar{\kup}}}  \per
\end{align}
For $\kup \approx \bar{\kup}$ this becomes $\theta_{k \bar{k}} < ( 2 / \kap)^{2/3} (\kup L)^{-1/3}$, which agrees with a similar estimate in \rref{Srednicki:1986xg}.  

From the arguments above, we obtain the cusp integral to be 
\begin{align}\label{eq:Hquad_Icusp}
	& \bigl| \mathcal{I}^{\rm (cusp)} (k + \bar{k}) \bigr|^2 = \Scusp \frac{L^{4/3}}{(\kup + \bar{\kup})^{8/3}} 
	\Com
	\kap \, m_H \sqrt{m_H L} < \kup , \bar{\kup} < M \sqrt{ML}
	\ , \nn
	& \qquad 
	\theta_{k \bar{k}} < \kap^{-2/3} \frac{(\kup + \bar{\kup})^{2/3} L^{-1/3} }{ \sqrt{ \kup \bar{\kup} }} \, {\rm (cone)}
	\Com
	\theta_{k+\bar{k}} < \Theta (\kup + \bar{\kup})^{-1/3}  L^{-1/3} \, {\rm (cone)}
\end{align}
with $0.2 \lesssim \Scusp \lesssim 10$.  
We have also used $\theta_{k \bar{k}} \ll 1$ to approximate $| {\bf k} + \bar{\bf k}| \approx \kup + \bar{\kup}$.  
Inserting \eref{eq:Hquad_Icusp} into \eref{eq:dNHH_of_I} we obtain the spectrum 
\begin{align}\label{eq:dNHH_cusp_1}
	dN_{HH}^{\rm (cusp)} &= \frac{(\gHHstr)^2}{4 (2\pi)^6} \Scusp \frac{L^{4/3} \kup \, \bar{\kup}}{(\kup + \bar{\kup})^{8/3}} d\kup \, d\Omega \, d\bar{\kup} \, d\bar{\Omega}
	\Com
	\kap \, m_H \sqrt{m_H L} < \kup , \bar{\kup} < M \sqrt{ML}
	\ , \nn
	& \qquad 
	\theta_{k \bar{k}} < \kap^{-2/3} \frac{(\kup + \bar{\kup})^{2/3} L^{-1/3} }{\sqrt{ \kup \bar{\kup}} } \, {\rm (cone)} 
	\Com
	\theta_{k+\bar{k}} < \Theta (\kup + \bar{\kup})^{-1/3}  L^{-1/3} \, {\rm (cone)} \per
\end{align}
The upper bound on $\theta_{k \bar{k}}$ implies that ${\bf k}$ and $\bar{\bf k}$ are approximately parallel to one another, and the upper bound on $\theta_{k+\bar{k}}$ implies that their sum points along the direction of the cusp.  
The geometry is such that the radiation is emitted into a pair of overlapping cones, and the angular integrations yield
\begin{align}\label{eq:Hquad_angle_int}
	\int d \Omega \, d \bar{\Omega} 
	\approx  \frac{(2\pi)^2}{4} \kap^{-4/3} \Theta^2 \frac{(\kup + \bar{\kup})^{2/3}}{\kup \bar{\kup}} L^{-4/3} \com
\end{align}
and the spectrum becomes
\begin{align}\label{eq:dNHH_cusp}
	dN_{HH}^{\rm (cusp)} &= \frac{(\gHHstr)^2}{16 (2\pi)^4} \kap^{-4/3} \Theta^2 \Scusp \frac{d\kup \, d\bar{\kup}}{(\kup + \bar{\kup})^{2}} 
	\Com
	\kap \, m_H \sqrt{m_H L} < \kup , \bar{\kup} < M \sqrt{ML} \per
\end{align}
The total energy emitted from a cusp is given by 
\begin{align}\label{eq:EHH_cusp}
	E_{HH}^{\rm (cusp)} = \int (\omega + \bar{\omega}) dN_{HH}^{\rm (cusp)} 
	\approx \frac{(\gHHstr)^2}{16 (2\pi)^4} \kap^{-4/3} \Theta^2 \Scusp M \sqrt{ML} \per
\end{align}
If the frequency of cusp appearance is $f_{\rm cusp} = f_c / T$ with $T = L /2$ is the loop oscillation period, then the average power emitted is
\begin{align}\label{eq:PHH_cusp}
	P_{HH}^{\rm (cusp)} = \GHHcusp \frac{( \gHHstr M)^2}{\sqrt{ML}}
\end{align}
where 
$\GHHcusp \equiv \frac{1}{8 (2\pi)^4} f_c \kap^{-4/3} \Theta^2 \, \Scusp$.  
Estimating $f_c \approx 1$ gives $10^{-5} < \GHHcusp < 10^{-2}$.  

Scalar boson pair radiation from a cusp has been calculated previously by \rref{Srednicki:1986xg}.  
Our calculation matches the UV-sensitive spectrum, \eref{eq:dNHH_cusp}, of the earlier reference.  
In calculating the total power, we integrate up to an energy of $M \sqrt{ML}$ where $1 / M$ is the string thickness and $\sqrt{ML}$ is the boost factor that translates between the cusp and loop rest frames (see \sref{sub:HiggsCusp}).  
This boost factor was overlooked in the previous calculations, and the power was found to be $O(M / L)$, typically a significant underestimate compared to \eref{eq:PHH_cusp}.

\addtocontents{toc}{\SkipTocEntry}
\subsubsection{Higgs-Higgs Emission from a Kink}\label{sub:HiggsQuadKink}

We calculate the spectrum of Higgs boson radiation from the kink by evaluating the spectrum in \eref{eq:dNHH_of_I} using the scalar integral in \eref{eq:I_scalar_kink}.  
After also generalizing the saddle point criterion, as discussed in \sref{sub:HiggsQuadCusp}, we obtain 
\begin{align}\label{eq:dNHH_kink_1}
	dN_{HH}^{\rm (kink)} &= \frac{(\gHHstr)^2}{4(2\pi)^6} \Skink \frac{L^{4/3} \, \kup \, \bar{\kup}}{(\kup + \bar{\kup})^{8/3}}  d\kup \, d\Omega \, d\bar{\kup} \, d\bar{\Omega} 
	\Com
	\kap \, m_H \sqrt{m_H L} < \kup , \bar{\kup} < M 
	\ , \nn
	& \qquad 
	\theta_{k \bar{k}} < \kap^{-2/3} \frac{(\kup + \bar{\kup})^{2/3} L^{-1/3} }{ \sqrt{ \kup \bar{\kup} } } \, {\rm (cone)}
	\Com
	\theta_{k+\bar{k}} < \Theta \, (\kup + \bar{\kup})^{-1/3} L^{-1/3}  \, {\rm (band)}
\end{align}
where $0.05 < \Skink < 10$.  
The momenta ${\bf k}$ and $\bar{\bf k}$ are separated by an angle $\theta_{k\bar{k}}$, and their sum is oriented in a band of angular with $\Theta (\kup + \bar{\kup})^{-1/3} L^{-1/3}$.  
Performing the angular integrations we obtain 
\begin{align}\label{eq:dNHH_kink}
	dN_{HH}^{\rm (kink)} &= \frac{(\gHHstr)^2}{8(2\pi)^4} \kap^{-4/3} \Theta \, \Skink L^{1/3} \frac{d\kup \, d\bar{\kup}}{(\kup + \bar{\kup})^{5/3}} 
	\Com
	\kap \, m_H \sqrt{m_H L} < \kup , \bar{\kup} < M \per
\end{align}
The spectrum is UV-sensitive, which allows us to neglect the lower limit, and upon integrating we find the total energy output to be 
\begin{align}\label{eq:EHH_kink}
	E_{HH}^{\rm (kink)} 
	= \int (\omega + \bar{\omega}) \, dN_{HH}^{\rm (kink)}
	\approx \frac{9 (\gHHstr)^2}{16(2\pi)^4} \kap^{-4/3} \Theta \, \Skink L^{1/3} M^{4/3} 
	\left( 1 - 5 \frac{\psi m_H \sqrt{m_H L} }{M } \right)
\end{align}
where we have used $4 / [ 3(2^{1/3}-1) ] \approx 5$ in the second term.  
If the loop contains $N_k$ kinks, then the average power output during one loop oscillation period ($T = L /2$) is given by 
\begin{align}\label{eq:PHH_kink}
	P_{HH}^{\rm (kink)} 
	= \GHHkink \frac{( \gHHstr M)^{2}}{(ML)^{2/3}} 
	\left( 1 - 5 \psi \frac{m_H \sqrt{m_H L} }{M } \right)
	\per
\end{align}
where 
$\GHHkink \equiv \frac{9}{8(2\pi)^4} N_{k} \kap^{-4/3}   \Theta \, \Skink$.  
Estimating $N_k \approx 1$ and using the range for $\Skink$ given above, the dimensionless prefactor can be estimated as $10^{-4} < \GHHkink < 10^{-1}$.

\addtocontents{toc}{\SkipTocEntry}
\subsubsection{Higgs-Higgs Emission from a Kink-Kink Collision}\label{sub:HiggsQuadKK}

To calculate the spectrum of Higgs boson radiation from a kink-kink collision we use the scalar integral from \eref{eq:I_scalar_kk} in the spectrum from \eref{eq:dNHH_of_I} to obtain 
\begin{align}\label{eq:dNHH_kk_1}
	dN_{HH}^{\rm (k-k)} &= \frac{(\gHHstr)^2}{4 (2\pi)^6} \Skk \frac{\kup \, \bar{\kup}}{(\omega + \bar{\omega})^4} d \omega \, d \Omega \, d \bar{\omega} \, d \bar{\Omega}
	\Com
	m_H < \omega , \bar{\omega} < M 
\end{align}
where $0.05 < \Skk < 200$.  
The radiation can be emitted isotropically; performing the angular integration gives a factor of $(4\pi)^2$ and leaves 
\begin{align}\label{eq:dNHH_kk}
	dN_{HH}^{\rm (k-k)} &= \frac{(\gHHstr)^2}{(2\pi)^4} \Skk \frac{\kup \, \bar{\kup}}{(\omega + \bar{\omega})^4} d \omega \, d \bar{\omega} 
	\Com
	m_H < \omega , \bar{\omega} < M \per
\end{align}
The total energy output of a kink-kink collision is calculated as 
\begin{align}\label{eq:EHH_kk}
	E_{HH}^{\rm (k-k)} 
	= \int (\omega + \bar{\omega}) \, d N_{HH}^{\rm (k-k)}
	= \frac{(\gHHstr)^2}{4 (2\pi)^4} \Skk M \per
\end{align}
If there are $N_{kk}$ kink-kink collisions during a loop oscillation period $T = L /2$ then the average power is found to be 
\begin{align}\label{eq:PHH_kk}
	P_{HH}^{\rm (k-k)} 
	= \GHHkk \frac{(\gHHstr M)^2}{ML} \per
\end{align}
where $\GHHkk \equiv \frac{1}{2 (2\pi)^4} N_{kk} \Skk$.  
For $N_{kk} \approx 1$ we can estimate $10^{-4} < \GHHkk < 10^{-1}$ using the range for $\Skk$ given above.  

\subsection{Z-Boson Emission via Linear Coupling}\label{sub:ZbosonEmission_Linear}

The interaction
\begin{equation}\label{eq:Sint_Z}
	\SZint = 
  \frac{\gZstr}{2} \left( \frac{\eta}{\sigma} \right)^2 \int d\sigma^{\mu\nu} Z_{\mu\nu}(\Xbb)
\end{equation}
allows Z bosons to be radiated from the string.  
The radiation calculation is carried out in \aref{app:Radiation}.  
The spectrum is given by \eref{eq:Zsingle_dN} after replacing $C = \gZstr (\eta / \sigma)^2$:  
\begin{align}\label{eq:dN_Z_final}
	dN_Z & = (\gZstr)^2 \left( \frac{\eta}{\sigma} \right)^4 \frac{\kup \, d\omega \, d\Omega}{2 (2\pi)^3} m_Z^2 \, \Pi(k) \per
\end{align}
In this expression $\omega = (\kup^2 + m_Z^2)^{1/2}$ with $m_Z$ the Z boson mass and $\Pi(k)$ is a functional of the stringworldsheet, given by \eref{eq:Pi_def}.  
In the following subsections we calculate the spectrum and total power in Z boson emission from cusps, kinks, and kink-kink collisions.

\addtocontents{toc}{\SkipTocEntry}
\subsubsection{Z Emission from a Cusp}\label{sub:ZbosonCusp}

The spectrum of Z boson emission from a cusp is calculated using \eref{eq:dN_Z_final} with the integral in \eref{eq:Pi_cusp}.  
Combining these formulae we obtain 
\begin{align}\label{eq:dNZ_cusp_1}
	dN_Z^{\rm (cusp)} & = 
	\frac{(\gZstr)^2}{2 (2\pi)^3} \left( \frac{\eta}{\sigma} \right)^4 \Tcusp \frac{L^{4/3}}{\kup^{5/3}} m_Z^2 d\kup \, d\Omega 
	\Com \nn
	& \qquad
	\kap \, m_Z \sqrt{m_Z L} < \kup < M \sqrt{ML} 
	\Com
	\theta < \Theta \, (\kup L)^{-1/3} \, {\rm (cone)}
	\per
\end{align}
where the dimensionless coefficient takes values $0.5 \lesssim \Tcusp \lesssim 50$.  
The direction of the outgoing Z boson lies within a cone centered at the cusp and has an opening angle $\Theta (\kup L)^{-1/3}$.  
We integrate over the solid angle to obtain the spectrum 
\begin{align}\label{eq:dNZ_cusp}
	dN_Z^{\rm (cusp)} & = 
	\frac{(\gZstr)^2}{4 (2\pi)^2} \Theta^{2} \left( \frac{\eta}{\sigma} \right)^4 \Tcusp m_Z^2 \frac{L^{2/3}}{\kup^{7/3}} d\kup
	\Com 
	\kap \, m_Z \sqrt{m_Z L} < \kup < M \sqrt{ML} \com
\end{align}
we integrate over the momentum to obtain the energy output from a single cusp 
\begin{align}\label{eq:EZ_cusp}
	E_Z^{\rm (cusp)} & 
	= \int \omega \, dN_{Z}^{\rm (cusp)} 
	= \frac{3(\gZstr)^2}{4 (2\pi)^2} \psi^{-1/3} \Theta^{2} \left( \frac{\eta}{\sigma} \right)^4 \Tcusp m_Z \sqrt{m_Z L}
\end{align}
and if cusps arise with a frequency $f_c / T$ where $T = L/2$ is the loop oscillation period, then the average power per loop oscillation is found to be 
\begin{align}\label{eq:PZ_cusp}
	P_{Z}^{\rm (cusp)} & 
	= \GZcusp \left( \frac{\eta}{\sigma} \right)^4 \frac{(\gZstr m_Z)^2}{\sqrt{m_Z L}} 
\end{align}
where the power coefficient is $\GZcusp \equiv \frac{3}{2 (2\pi)^2} \Tcusp f_c \psi^{-1/3} \Theta^{2}$.  
Assuming $f_c \approx 1$ we estimate $10^{-4} \lesssim \GZcusp \lesssim 10^{-1}$.

\addtocontents{toc}{\SkipTocEntry}
\subsubsection{Z Emission from a Kink}\label{sub:ZbosonKink}

To calculate the spectrum of Z boson emission from a single kink, we use the expression \eref{eq:dN_Z_final} along with the expression \eref{eq:Pi_kink} for $\Pi(k)$ for a kink to find
\begin{align}\label{eq:dN_Z_kink_1}
	dN_Z^{\rm kink} & = \frac{(\gZstr)^2}{2(2\pi)^3} \left( \frac{\eta}{\sigma} \right)^4 \Tkink \frac{L^{2/3}}{\kup^{7/3}} \, m_Z^2 \, d\kup d\Omega
	\Com \nn
	& \qquad
	\kap \, m_Z \sqrt{m_Z L} < \kup < M
	\, , \quad
	\theta < \Theta \, (\kup L)^{-1/3} \, {\rm (band)}
\end{align}
where $0.5 < \Tkink < 100$ 
Radiation is emitted in a band with angular width $\Theta (\kup L)^{-1/3}$, and we integrate over the solid angle to find 
\begin{align}\label{eq:dN_Z_kink}
	dN_Z^{\rm kink} & = \frac{(\gZstr)^2}{2(2\pi)^2} \Theta \left( \frac{\eta}{\sigma} \right)^4 \Tkink \, \frac{L^{1/3}}{\kup^{8/3}} \, m_Z^2 \, d\kup
	\Com
	\kap \, m_Z \sqrt{m_Z L} < \kup < M
\end{align}
The total energy emitted by a kink during one loop oscillation is
\begin{align}\label{eq:EZ_kink}
	E_Z^{\rm (kink)} 
	= \int \omega \, dN_Z^{\rm kink}
	= \frac{3(\gZstr)^2}{4(2\pi)^2} \psi^{-2/3} \Theta \left( \frac{\eta}{\sigma} \right)^4 \Tkink m_Z 
	\left( 1 - \psi^{2/3} \frac{ m_Z L^{1/3}}{M^{2/3}} \right)
	\com
\end{align}
and if there are $N_k$ kinks on the loop then the average radiated power during one loop oscillation period ($T = L/2$) is 
\begin{align}\label{eq:PZ_kink}
	P_Z^{\rm (kink)} = \GZkink \left( \frac{\eta}{\sigma} \right)^4 \, \frac{(\gZstr m_Z)^2}{m_Z L}
	\left( 1 - \psi^{2/3} \frac{ m_Z L^{1/3}}{M^{2/3}} \right)
\end{align}
with $\GZkink = \frac{3}{2(2\pi)^2} \psi^{-2/3} \, \Theta \Tkink \, N_k$.  
Estimating $N_k \approx 1$ gives $10^{-2} < \GZkink < 10$.  

\addtocontents{toc}{\SkipTocEntry}
\subsubsection{Z Emission from a Kink-Kink Collision}\label{sub:ZbosonKinkKink}

Inserting \eref{eq:Pi_kk} into \eref{eq:dN_Z_final} we obtain the spectrum of Z boson emission from a collision of kinks to be 
\begin{align}\label{eq:dNZ_kk_1}
	dN_Z^{\rm (k-k)} & = \frac{(\gZstr)^2}{2 (2\pi)^3} \left( \frac{\eta}{\sigma} \right)^4 \Tkk \frac{\kup}{\omega^4} m_Z^2  \, d\omega \, d\Omega
	\Com  
	m_Z < \omega < M 
\end{align}
with the constant $0.1 < \Tkk < 50$. 
The emission is isotropic, and after performing the angular integration we obtain 
\begin{align}\label{eq:dNZ_kk}
	dN_Z^{\rm (k-k)} \approx \frac{(\gZstr)^2}{4 (2\pi)^2} \left( \frac{\eta}{\sigma} \right)^4 \Tkk \frac{\kup}{\omega^4} m_Z^2 \, d\omega 
	\Com  
	m_Z < \omega < M 	
\end{align}
The total energy emitted by a kink-kink collision is found to be 
\begin{align}\label{eq:EZ_kk}
	E_Z^{\rm (k-k)}  
	= \int \omega \, dN_Z^{\rm (k-k)}
	= \frac{(\gZstr)^2}{4 (2\pi)^2} \left( \frac{\eta}{\sigma} \right)^4 \Tkk m_Z \per
\end{align}
If $N_{kk}$ such collisions occur during one loop oscillation period, $T = L /2$, then the average power is 
\begin{align}\label{eq:PZ_kk}
	P_Z^{\rm (k-k)} = \GZkk  \, \left( \frac{\eta}{\sigma} \right)^4 \frac{(\gZstr m_Z)^2}{m_Z L}
\end{align}
with $\GZkk \equiv \frac{1}{2(2\pi)^2} N_{kk} \, \Tkk$.
Estimating $N_{kk} \approx 1$ gives $10^{-3} < \GHkk < 1$.

\subsection{Z Boson Emission via Quadratic Coupling}\label{sub:ZbosonEmission_Quadratic}

An interaction of the form 
\begin{equation}\label{eq:Sint_ZZ}
	\SZZint = \gZZstr \left( \frac{\eta}{\sigma} \right)^4 \int d^2\sigma \, \sqrt{-\gamma} \, Z_{\mu}(\Xbb) Z^{\mu}(\Xbb)
\end{equation}
also allows Z bosons to be radiated from the string.  
For heavy strings, the coefficient $(\eta/\sigma)^4$ is very small, and this radiation channel is negligible.  
However, we present the calculation of the radiation spectra for completeness.  
The spectrum is given by \eref{eq:Zquad_dN} after replacing $C = \gZZstr (\eta / \sigma)^4$,
\begin{align}\label{eq:dNZZ_of_I}
	dN_{ZZ} = 4 (\gZZstr)^2 \left( \frac{\eta}{\sigma} \right)^{8} \frac{\kup \, d\omega \, d\Omega}{2(2\pi)^3} \, \frac{\bar{\kup} \, d\bar{\omega} \, d\bar{\Omega}}{2(2\pi)^3} \, | \mathcal{I}(k + \bar{k}) |^2 \ 
	\com
\end{align}
where $k^{\mu} = \bigl\{ \omega \, , \, {\bf k} \bigr\}$ and $\omega = ( m_Z^2 + \kup^2 )^{1/2}$
with similar definitions for the barred quantities.  
Note the similarity between \eref{eq:dNZZ_of_I} and the spectrum of Higgs boson pair radiation given by \eref{eq:dNHH_of_I}.  
Since both spectra depend on the same scalar integral, $\Ical(k + \bar{k})$, we can simply carry over all the results from \sref{sub:HiggsPairEmission_Quadratic}.  
We need only to make the replacement $(\gHHstr)^2 \to 4 (\gZZstr)^2 (\eta / \sigma)^8$.

\addtocontents{toc}{\SkipTocEntry}
\subsubsection{Z-Z Emission from a Cusp}\label{sub:ZQuadCusp}

We calculate the spectrum of Z boson radiation from a cusp following \sref{sub:HiggsQuadCusp}.  
We find the spectrum
\begin{align}\label{eq:dNZZ_cusp}
	&dN_{ZZ}^{\rm (cusp)} = \frac{(\gZZstr)^2}{4(2\pi)^4} \kap^{-4/3}  \Theta^2 \left( \frac{\eta}{\sigma} \right)^{8} \Scusp
	\frac{d \kup \, d\bar{\kup}}{(\kup + \bar{\kup})^{2}}
	\Com 
	\kap \, m_Z \sqrt{m_Z L} < \kup , \bar{\kup} < M \sqrt{ML} \com
\end{align}
the energy radiated per cusp event
\begin{align}\label{eq:EZZ_cusp}
	E_{ZZ}^{\rm (cusp)} 
	= \frac{(\gZZstr)^2}{4(2\pi)^4} \kap^{-4/3}  \Theta^2 \left( \frac{\eta}{\sigma} \right)^{8} \Scusp M \sqrt{ML} \com
\end{align}
and the average power output if cusps arise with frequency $2 f_c / L$
\begin{align}\label{eq:PZZ_cusp}
	P_{ZZ}^{\rm (cusp)} = \GZZcusp \left( \frac{\eta}{\sigma} \right)^{8} \frac{(\gZZstr M)^2}{\sqrt{ML}} \per
\end{align}
The dimensionless coefficient is defined as 
$\GZZcusp \equiv \frac{1}{2(2\pi)^4} f_c \kap^{-4/3}  \Theta^2 \Scusp$ 
and it may be estimated as $10^{-4} < \GZZcusp < 10^{-2}$.  

\addtocontents{toc}{\SkipTocEntry}
\subsubsection{Z-Z Emission from a Kink}\label{sub:ZQuadKink}

We calculate the spectrum of Z boson radiation from a kink following \sref{sub:HiggsQuadKink}.  
We find the spectrum
\begin{align}\label{eq:dNZZ_kink}
	&dN_{ZZ}^{\rm (kink)} = \frac{(\gZZstr)^2}{2(2\pi)^4} \kap^{-4/3} \Theta \left( \frac{\eta}{\sigma} \right)^{8} \Skink L^{1/3}
	\frac{ d \kup \, d\bar{\kup} }{(\kup + \bar{\kup})^{5/3}} 
	\Com 
	\kap \, m_Z \sqrt{m_Z L} < \kup , \bar{\kup} < M \com
\end{align}
the energy radiated per kink during one loop oscillation 
\begin{align}\label{eq:EZZ_kink}
	E_{ZZ}^{\rm (kink)} 
	\approx \frac{9(\gZZstr)^2}{4(2\pi)^4} \kap^{-4/3} \Theta \left( \frac{\eta}{\sigma} \right)^{8} \Skink L^{1/3} M^{4/3} 
	\left( 1 - 5 \psi \frac{m_Z \sqrt{m_Z L} }{M } \right)
	\com
\end{align}
and the average power emitted from a loop containing $N_k$ kinks 
\begin{align}\label{eq:PZZ_kink}
	P_{ZZ}^{\rm (kink)} = \GZZcusp \left( \frac{\eta}{\sigma} \right)^{8} \frac{(\gZZstr M)^2}{(ML)^{2/3}} 
	\left( 1 - 5 \psi \frac{m_Z \sqrt{m_Z L} }{M } \right)
	\per
\end{align}
The dimensionless coefficient is defined by 
$\GZZkink \equiv \frac{9}{2(2\pi)^4} N_{k} \kap^{-4/3}   \Theta \, \Skink$ 
 and it can be estimated as $10^{-3} < \GZZkink < 10^{-1}$.  

\addtocontents{toc}{\SkipTocEntry}
\subsubsection{Z-Z Emission from a Kink-Kink Collision}\label{sub:ZQuadKK}

We calculate the spectrum of Z boson radiation from a collision of two kinks following \sref{sub:HiggsQuadKink}.  
We find the spectrum
\begin{align}\label{eq:dNZZ_kk}
	dN_{ZZ}^{\rm (k-k)} &= \frac{(\gZZstr)^2}{4 (2\pi)^4} \left( \frac{\eta}{\sigma} \right)^{8} \Skk \frac{\kup \, \bar{\kup}}{(\omega + \bar{\omega})^4} d \omega \, d \bar{\omega} 
	\Com
	m_Z < \omega , \bar{\omega} < M \com
\end{align}
the energy radiated during the collision
\begin{align}\label{eq:EZZ_kk}
	E_{ZZ}^{\rm (k-k)} 
	= \frac{(\gZZstr)^2}{16 (2\pi)^4} \left( \frac{\eta}{\sigma} \right)^{8} \Skk M \com
\end{align}
and the average power radiated from a loop that experiences $N_{kk}$ collisions during one loop oscillation period
\begin{align}\label{eq:PZZ_kk}
	P_{ZZ}^{\rm (k-k)} 
	= \GZZkk \left( \frac{\eta}{\sigma} \right)^{8} \frac{(\gZZstr M)^2}{ML} \per
\end{align}
The dimensionless coefficient is defined by $\GZZkk \equiv \frac{1}{8 (2\pi)^4} N_{kk} \Skk$, and we can estimate $10^{-5} < \GZZkk < 10^{-2}$.


\subsection{Fermion Emission via Aharonov-Bohm Coupling}\label{sub:ABEmission}

The cosmic string can radiate fermions through a direct coupling, such as the ones we have been studying for the Higgs and Z bosons, or through a non-local AB interaction.  
SM fermions couple directly to the string worldsheet through interactions of the form 
\begin{align}
	S_{\rm int}^{(\psi)} = \frac{\gpsistr}{M} \left( \frac{\eta}{\sigma} \right)^2 \int d^2 \sigma \, \sqrt{-\gamma} \, \bar{\Psi}(\Xbb^{\mu}) \Psi(\Xbb^{\mu}) 
\end{align}
where $\gpsistr$ is a dimensionless coupling constant, and the factor of $(\eta/\sigma)^2$ arises from the mixing between the Higgs field and the HS scalar field \cite{Hyde:2013fia}.  
Note that dimensional analysis requires the string mass scale to appear in the denominator.  
The radiation calculation with $S_{\rm int}^{(\psi)}$ is very similar to the case of Higgs radiation via the quadratic interaction, see \sref{sub:SpinorRad_1}.  
We find the spectrum of $\psi$ radiation to be 
\begin{align}
	dN_{\psi \psi} = 4 \left( \frac{\eta}{\sigma} \right)^4 \left( \frac{\gpsistr}{\gHHstr} \right)^2 \frac{k \cdot \bar{k} - m_{\psi}^2}{M^2}dN_{HH}
\end{align}
where $dN_{HH}$ is the spectrum of Higgs radiation, given by \eref{eq:dNHH_of_I}.  
Because of the mixing angle factor, $(\eta / \sigma)^4 \ll 1$, this radiation channel is inefficient.

The non-local AB interaction provides an additional channel for particle production from the cosmic string \cite{Alford:1988sj}.  
Refs.~\cite{JonesSmith:2009ti,Chu:2010zzb,Steer:2010jk} studied the AB radiation of scalars, fermions, and vectors from a string. 
In these calculations, the authors assumed that the string carries only one kind of magnetic flux, which is usually the case.  
The structure of the dark string, however, is more complex.  
As we saw in \rref{Hyde:2013fia}, the string core contains flux of the HS $X^{\mu}$ field and the dressing contains flux of the SM $Z^{\mu}$ field.  
When a fermion travels around the perimeter of the string, outside of both the core and the dressing, its wavefunction picks up an AB phase due to both fluxes, and the overall phase is given by $2\pi \theta_q$, where $\theta_q$ is defined in \eref{thetaq}.  
On the other hand, when the fermion makes a loop around the core by passing through the region of space containing the dressing fields, it will acquire a different AB phase.  

In order to setup the radiation calculation we must know the effective AB interaction of the fermions with the string.  
The discussion above is intended to illustrate that this interaction will be scale dependent.  
At energies below the inverse dressing width, $\sim 1 / \eta$, the core plus dressing can be treated together as a zero width string.  
In this limit the structure of the string is unimportant, and the AB interaction can be derived following Refs.~\cite{JonesSmith:2009ti,Chu:2010zzb,Steer:2010jk} with the AB phase given by $\theta_{q}$.  
At higher energies the Compton wavelength of the radiation drops below the dressing thickness.  
Here the effective coupling will presumably decrease as the particle ``sees'' less and less of the flux carried by the dressing.  
This behavior is in contrast with the Higgs and Z boson radiation channels that we considered previously.  
In those cases, the light SM fields coupled directly to the string core itself, and the dressing was neglected.  

In light of the discussion above, we will proceed as follows.  
We calculate the spectrum of radiation due to the AB interaction where the coupling is set by the AB phase $\theta_q$.  
If the thickness of the string dressing is $\sim 1 / \eta$, then this spectrum is valid up to energies $\kup \approx \eta \sqrt{\eta L}$ for the cusp or $\kup \approx \eta$ for the kink and kink collision.  
At higher energies, we suppose that the effective coupling begins to decrease as the fermion radiation begins to penetrate inside of the dressing, and consequently the spectrum drops sharply.

The AB interaction can be treated perturbatively as follows.  
Let $V_{\mu}(x)$ be the appropriate linear combination of the $X_{\mu}$ and $Z_{\mu}$ gauge fields that couples to the fermions, and let $g_{\psi}$ be the coupling constant.  
Then the interaction is given by 
\begin{align}\label{VPsiint}
	\Lcal_{\rm eff} & = g_{\psi} V_{\mu}(x) \bar{\Psi}(x) \gamma^{\mu} \Psi(x) \per
\end{align}
We treat $V_{\mu}$ as a classical background field induced by the flux that the string carries: $\Phi = (2\pi / g_{\psi})\theta_q$.  This lets us write (Lorentz gauge, $\partial_{\mu} V^{\mu} = 0$) \cite{Alford:1988sj} 
\begin{equation}
	V_{\mu} = - \frac{\Phi}{2} \int_{\rm ret.} \frac{d^4p}{(2\pi)^4} \frac{i p^{\nu}}{p^2} \int d\sigma_{\mu \nu} \, e^{-ip \cdot (x-\Xbb)} 
\end{equation}
where the integration contour extends above the poles at $p^0 = \pm \abs{\bf p}$, as in the calculation of a retarded Green's function.  
Note that $V_{\mu}(x)$ has support outside of the string, unlike the purely local interactions in \erefs{eq:Sint}{eq:Sintquadratic}.  

The interaction inEq.~(\ref{VPsiint}) allows the string to 
radiate fermion pairs with momenta $k^{\mu} = \bigl\{ \omega = \sqrt{m_{\psi}^2 + \kup^2} \, , \, {\bf k} \bigr\}$ and $\bar{k}^{\mu} = \bigl\{ \bar{\omega} = \sqrt{m_{\psi}^2 + \bar{\kup}^2} \, , \, \bar{\bf k} \bigr\}$.  
The spectrum is given by \eref{eq:Psi_dN_2} after replacing $C = - (2\pi \theta_{q})/2$:  
\begin{align}\label{eq:dNpsi_of_I}
	dN_{\psi \psi}
	& = \frac{(2\pi \theta_{q})^2}{8 (2\pi)^6} \, \Pi(k + \bar{k}) \, \kup d\omega d\Omega \bar{\kup} d\bar{\omega} d\bar{\Omega} 
\end{align}
where $\Pi$ is given by \eref{eq:Pi}.

\addtocontents{toc}{\SkipTocEntry}
\subsubsection{Fermion AB Emission from a Cusp}\label{sub:PsiQuadCusp}

We find the spectrum of radiation from a cusp by inserting \eref{eq:Pi_cusp} into \eref{eq:dNpsi_of_I}:  
\begin{align}
	dN_{\AB}^{\rm (cusp)}
	& = \frac{(2\pi\theta_{q})^2}{8 (2\pi)^6} \, \Tcusp \frac{L^{4/3} \kup \bar{\kup} }{(\kup + \bar{\kup})^{8/3}} \,  d\kup \, d\Omega \, d\bar{\kup} \, d\bar{\Omega} 
	\Com
	\psi \, m_{\psi} \sqrt{ m_{\psi} L } < \kup , \bar{\kup} < \eta \sqrt{\eta L} 
	\com \nn
	& \qquad
	\theta_{k \bar{k}} < \psi^{-2/3} \frac{ (\kup + \bar{\kup})^{2/3} L^{-1/3} }{ \kup^{1/2} \bar{\kup}^{1/2} } \, {\rm (cone)}
	\Com
	\theta_{k+\bar{k}} < \Theta (\kup + \bar{\kup})^{-1/3} L^{-1/3} \, {\rm (cone)}
\end{align}
where $0.5 \lesssim \Tcusp \lesssim 50$.  
Recall from the discussion of \sref{sub:HiggsQuadCusp} that the momentum sum ${\bf k} + \bar{\bf k}$ is oriented within a cone of angle $\Theta (\kup + \bar{\kup})^{-1/3} L^{-1/3}$ centered on the cusp, and the angle between ${\bf k}$ and $\bar{\bf k}$ cannot exceed $\psi^{-2/3} (\kup + \bar{\kup})^{2/3} L^{-1/3} / \sqrt{ \kup \bar{\kup} }$.  
Upon performing the angular integrations as in \eref{eq:Hquad_angle_int}, we obtain 
\begin{align}
	dN_{\AB}^{\rm (cusp)}
	& = \frac{(2\pi\theta_{q})^2}{32 (2\pi)^4} \psi^{-4/3} \Theta^{2} \Tcusp  \, \frac{d\kup d\bar{\kup}}{(\kup + \bar{\kup})^{2}} 
	\Com
	\psi \, m_{\psi} \sqrt{ m_{\psi} L } < \kup , \bar{\kup} < \eta \sqrt{\eta L} 
\end{align}
We calculate the total energy output as 
\begin{align}
	E_{\AB}^{\rm (cusp)} = \int (\omega + \bar{\omega}) \, dN_{\AB}^{\rm (cusp)}
	\approx \frac{(2\pi\theta_{q})^2}{32 (2\pi)^4} (\psi^{-4/3} \Theta^{2} \Tcusp m_{\psi} \eta  \sqrt{\eta L}
\end{align}
and the average power output per loop oscillation as 
\begin{align}
	P_{\AB}^{\rm (cusp)} = \GABcusp \frac{(2\pi\theta_{q} \eta)^2}{\sqrt{\eta L}}
\end{align}
where 
$\GABcusp \equiv \frac{1}{32 (2\pi)^4} \psi^{-4/3} \Theta^{2} f_c \Tcusp $.  
Using the range for $\Tcusp$ given above, we can estimate $10^{-5} \lesssim \GABcusp \lesssim 10^{-2}$.

\addtocontents{toc}{\SkipTocEntry}
\subsubsection{Fermion AB Emission from a Kink}\label{sub:PsiQuadKink}

We find the spectrum of radiation from a kink by inserting \eref{eq:Pi_kink} into \eref{eq:dNpsi_of_I}:  
\begin{align}
	dN_{\AB}^{\rm (kink)}
	& = \frac{(2\pi\theta_{q})^2}{8 (2\pi)^6} \, \Tkink \frac{L^{2/3}}{( \kup + \bar{\kup})^{10/3}}\, \kup \bar{\kup} \, d\kup d\Omega d\bar{\kup} d\bar{\Omega} 
	\Com
	\psi \, m_{\psi} \sqrt{ m_{\psi} L } < \kup , \bar{\kup} < \eta
	\com \nn
	& \qquad
	\theta_{k \bar{k}} < \psi^{-2/3} \frac{ (\kup + \bar{\kup})^{2/3} L^{-1/3} }{ \kup^{1/2} \bar{\kup}^{1/2} }
	\Com
	\theta_{k+\bar{k}} < \Theta ( \kup + \bar{\kup})^{-1/3} L^{-1/3}
\end{align}
where $0.5 \lesssim \Tkink \lesssim 100$.  
Recall that ${\bf k} + \bar{\bf k}$ is oriented in a ribbon with angular width $\theta_{k + \bar{k}}$, and the opening angle between ${\bf k}$ and $\bar{\bf k}$ does not exceed $\theta_{k \bar{k}}$.  
After performing the angular integrations we obtain
\begin{align}
	dN_{\AB}^{\rm (kink)}
	& = \frac{(2\pi\theta_{q})^2}{16 (2\pi)^4} \, \psi^{-4/3} \Theta \, \Tkink \frac{ d\kup d\bar{\kup} }{ (\kup + \bar{\kup})^{7/3} L^{1/3} }
	\Com
	\psi \, m_{\psi} \sqrt{ m_{\psi} L } < \kup , \bar{\kup} < \eta \per
\end{align}
We calculate the total energy output as 
\begin{align}
	E_{\AB}^{\rm (kink)} = \int (\omega + \bar{\omega}) \, dN_{\AB}^{\rm (kink)}
	\approx \frac{9  (2\pi\theta_{q})^2}{16 (2\pi)^4} \, \psi^{-4/3} \Theta \, \Tkink \frac{ \eta^{2/3} }{ L^{1/3} } 
	\left( 1 - \psi^{2/3} \frac{m_{\psi} L^{1/3}}{\eta^{2/3}}  \right)
	\com
\end{align}
and the average power output from $N_k$ kinks during one loop oscillation period ($T = L /2$) as
\begin{align}
	P_{\AB}^{\rm (kink)} = \GABkink \frac{(2\pi\theta_{q} \eta)^2}{ (\eta L)^{4/3} }
	\left( 1 - \psi^{2/3} \frac{m_{\psi} L^{1/3}}{\eta^{2/3}}  \right)
\end{align}
where 
$\GABkink \equiv \frac{9}{8 (2\pi)^4} \psi^{-4/3} \Theta \Tkink N_{k}$.  
Using the range for $\Tkink$ given above along with $N_{k} \approx 1$, we can estimate $10^{-2} \lesssim \GABkink \lesssim 1$.

\addtocontents{toc}{\SkipTocEntry}
\subsubsection{Fermion AB Emission from a Kink-Kink Collision}\label{sub:PsiQuadKK}

We find the spectrum radiation from a kink collision by inserting \eref{eq:Pi_kk} into \eref{eq:dNpsi_of_I}:  
\begin{align}
	dN_{\AB}^{\rm (k-k)}
	& = \frac{(2\pi\theta_{q})^2}{8 (2\pi)^6} \, \Tkk \frac{1}{(\omega + \bar{\omega})^2} \frac{1}{(k + \bar{k})^2} \, \kup d\omega d\Omega \bar{\kup} d\bar{\omega} d\bar{\Omega} 
	\Com m_{\psi} < \omega, \bar{\omega} < \eta 
\end{align}
with $0.1 < \Tkk < 50$.  
In this case, the emission is isotropic, and we can estimate $(k + \bar{k})^2 \approx 2 \omega \bar{\omega}$ up to an $O(1)$ factor associated with the angle between ${\bf k}$ and $\bar{\bf k}$.  
The angular integration is trivial, and we find 
\begin{align}
	dN_{\AB}^{\rm (k-k)}
	& = \frac{(2\pi\theta_{q})^2}{64 (2\pi)^4} \, \Tkk \frac{1}{(\omega + \bar{\omega})^2} d\omega d\bar{\omega} 
	\Com m_{\psi} < \omega, \bar{\omega} < \eta \per
\end{align}
We calculate the total energy radiated as 
\begin{align}
	E_{\AB}^{\rm (k-k)} = \int (\omega + \bar{\omega}) \, dN_{\AB}^{\rm (k-k)} 
	= \frac{(2\pi\theta_{q})^2}{64 (2\pi)^4} \Tkk \, \eta \com
\end{align}
and the average power emitted from a loop which experiences $N_{kk}$ collisions during a loop 
oscillation period ($T = L/2$) is found to be 
\begin{align}
	P_{\AB}^{\rm (k-k)} = \GABkk \frac{(2\pi\theta_{q} \eta)^2}{\eta L}
\end{align}
where 
$\GABkk \equiv \frac{1}{ 32 (2\pi)^4 } \, N_{kk} \, \Tkk$.  
Using the parameter ranges given above along with $N_{kk} \approx 1$, we can estimate $10^{-6} < \GABkk < 10^{-3}$.

\section{Scattering Cross Sections}
\label{abscattering}

The interactions discussed in \sref{sec:Introduction} allow SM particles to scatter off of the dark string.  
Interactions of the Higgs and Z bosons with the string, given by \erefs{eq:Sint}{eq:Sintquadratic}, will lead to a ``hard core'' scattering, and the AB phases of the SM fermions, given by \eref{thetaq}, will lead to a non-local AB scattering.  
If the couplings are comparable for the direct and the AB interactions, then the latter generally dominates \cite{VilenkinShellard:1994}, and therefore we focus on AB scattering here.  
Moreover, in the cosmological context the dark string will scatter from the SM plasma, which consists mostly of electrons and nuclei at late times.  

The AB interaction allows fermions to scatter from a cosmic string.  
The scattering cross section (per length of string) was found to be \cite{Alford:1988sj} 
\begin{equation}
\frac{d\sigma_{\rm AB}}{d\theta} = \frac{\sin^2(\pi\theta_q)}{2\pi k_\perp \sin^2(\theta/2)}
\end{equation}
where the AB phase for SM fermions, $\theta_q$,  is given in \eref{thetaq}, and
$k_\perp$ is the magnitude of the momentum transverse to the string.  
Inserting the expression for $\theta_q$ and expanding in the $\theta_q \ll 1$ limit gives
\begin{equation}
	\frac{d\sigma_{\rm AB}}{d\theta} \approx \frac{2\pi \cos^2 \theta_W \sin^2 \epsilon}{\gX^2 \, k_\perp \sin^2(\theta/2)} q^2
\end{equation}
where $q$ is the electromagnetic charge of the fermion.

To study the motion of strings through the cosmological medium, we are interested in the drag (momentum transfer) experienced by the string.  
This is calculated in terms of a ``transport cross section'' (see \cite{VilenkinShellard:1994}) given by
\begin{align}\label{eq:sigt_AB}
	\sigma_{\rm AB,t}({\bf k}) = \int_0^{2\pi} d\theta \frac{d\sigma_{\rm AB}}{d\theta} (1-\cos\theta)
	= \frac{2}{k_\perp}\sin^2(\pi\theta_q) 
	\approx \frac{8\pi^2 \cos^2 \theta_W \sin^2 \epsilon}{\gX^2} \frac{q^2}{k_\perp} \per
\end{align}
To obtain the total drag due to the entire medium, we must sum over the various species with their respective charges $q$.  

The derivation of the AB phase, given by \eref{thetaq}, assumed that the particle circumnavigates the string on a length scale larger than the width of the SM dressing.  
In this way, the particle trajectory encloses both the flux carried by the thin HS string core and the thick SM dressing.  
This length scale is microscopic, $\Delta x \sim \eta^{-1} \approx 10^{-16} \cm$, and therefore this assumption is well-justified for the cosmological medium at late times, where the inter-particle spacing is much larger than $\Delta x$.  

\section{Summary and Conclusion}\label{sec:Conclusion}

The dark string couples to the SM fields through the local interactions in \erefs{eq:Sint}{eq:Sintquadratic} and through the non-local Aharonov-Bohm interactions of charged fermions.  
These interactions lead to radiation of Higgs bosons, Z bosons, and fermions from cusps, kinks, and kink collisions on cosmic strings.  
The total power emitted in each of various channels is summarized as follows.  
For Higgs emission via a linear coupling 
\begin{subequations}
\begin{align}
	P_{H}^{\rm (cusp)} & = \Gamma_{H}^{\rm (cusp)} \frac{(\gHstr \eta )^2}{\sqrt{m_H L}} \label{eq:IV1a}
	&&
	10^{-4} < \Gamma_{H}^{\rm (cusp)} < 10^{-1}
	\\
	P_{H}^{\rm (kink)}  & =  \GHkink \frac{(\gHstr\eta)^2}{m_H L} \left( 1 - \psi^{2/3} \frac{m_H L^{1/3}}{M^{2/3}} \right)
	&&
	10^{-3} < \GHkink < 1
	\\
	P_{H}^{\rm (k-k)} &=  \GHkk \frac{(\gHstr\eta)^2}{m_H L}
	&&
	10^{-2} < \GHkk < 10
	\com
\end{align}
\end{subequations}
for Higgs emission via a quadratic coupling 
\begin{subequations}
\begin{align}
	P_{HH}^{\rm (cusp)} & = \GHHcusp \frac{ (\gHHstr M)^2 }{\sqrt{ML}} 
	&&
	10^{-5} < \GHHcusp < 10^{-2} \label{eq:dom_channel}
	\\
	P_{HH}^{\rm (kink)}  & =  \GHHkink \frac{(\gHHstr M)^{2}}{(ML)^{2/3}} \left( 1 - 5 \psi \frac{m_H \sqrt{m_H L} }{M } \right)
	&&
	10^{-4} < \GHHkink < 10^{-1}
	\\
	P_{HH}^{\rm (k-k)} &=  \GHHkk \frac{(\gHHstr M)^2}{ML} 
	&&
	10^{-4} < \GHHkk < 10^{-1} \label{eq:IV2c}
	\com
\end{align}
\end{subequations}
for Z boson emission via a linear coupling 
\begin{subequations}
 \begin{align}
	 P_{Z}^{\rm (cusp)} & = \GZcusp \left( \frac{\eta}{\sigma} \right)^4 \frac{(\gZstr m_Z)^2}{\sqrt{m_Z L}} 
	&&
	10^{-4} < \GZcusp < 10^{-1}
	\\
	P_Z^{\rm (kink)} & =  \GZkink \left( \frac{\eta}{\sigma} \right)^4 \frac{(\gZstr m_Z)^2}{m_Z L} \left( 1 - \psi^{2/3} \frac{ m_Z L^{1/3}}{M^{2/3}} \right)
	&&
	10^{-2} < \GZkink < 10
	\\
	P_Z^{\rm (k-k)} & = \GZkk  \, \left( \frac{\eta}{\sigma} \right)^4 \frac{(\gZstr m_Z)^2}{m_Z L}	&&
	10^{-3} < \GZkk < 1
	\com
\end{align}
\end{subequations}
for Z boson emission via a quadratic coupling 
\begin{subequations}
\begin{align}
	P_{ZZ}^{\rm (cusp)} & = \GZZcusp \left( \frac{\eta}{\sigma} \right)^{8} \frac{ (\gZZstr M)^2 }{\sqrt{ML}} \label{PZZcusp_summary}
	&&
	10^{-4} < \GZZcusp < 10^{-2}
	\\
	P_{ZZ}^{\rm (kink)}  & =  \GZZkink \left( \frac{\eta}{\sigma} \right)^{8} \frac{(\gZZstr M)^{2}}{(ML)^{2/3}} \left( 1 - 5 \psi \frac{m_Z \sqrt{m_Z L} }{M } \right)
	&&
	10^{-3} < \GZZkink < 10^{-1}
	\\
	P_{ZZ}^{\rm (k-k)} &=  \GZZkk \left( \frac{\eta}{\sigma} \right)^{8} \frac{(\gZZstr M)^2}{ML} 
	&&
	10^{-5} < \GZZkk < 10^{-2} \com
\end{align}
\end{subequations}
and for fermion emission via the AB interaction 
\begin{subequations}
\begin{align}
	P_{\AB}^{\rm (cusp)} &= \GABcusp \frac{(2\pi\theta_{q} \eta)^2}{\sqrt{\eta L}}
	&&
	10^{-5} < \GABcusp < 10^{-2}
	\\
	P_{\AB}^{\rm (kink)} &= \GABkink \frac{(2\pi\theta_{q} \eta)^2}{(\eta L)^{4/3}} 
	\left( 1 - \psi^{2/3} \frac{m_{\psi} L^{1/3}}{\eta^{2/3}}  \right)
	&&
	10^{-2} < \GABkink < 1
	\\
	P_{\AB}^{\rm (k-k)} &= \GABkk \frac{(2\pi\theta_{q} \eta)^2}{\eta L} 
	&&
	10^{-6} < \GABkk < 10^{-3} \per
\end{align}
\end{subequations}
Here $\kap \approx 0.1$ [see \eref{eq:kmin_bound}] and the other dimensionless coefficients ($\Gamma$ factors) depend on undetermined parameters that characterize the radiating string segment, \eg, the curvature nearby to the cusp or the sharpness of the kink.  
We quantify our ignorance of these parameters, as described in \aref{app:ST_integrals}, and this leads to the ranges shown above.  
The kink expressions are only valid for small $L$ where the power is positive.  

\begin{figure}[t]
\begin{center}
\includegraphics[width=0.33\textwidth]{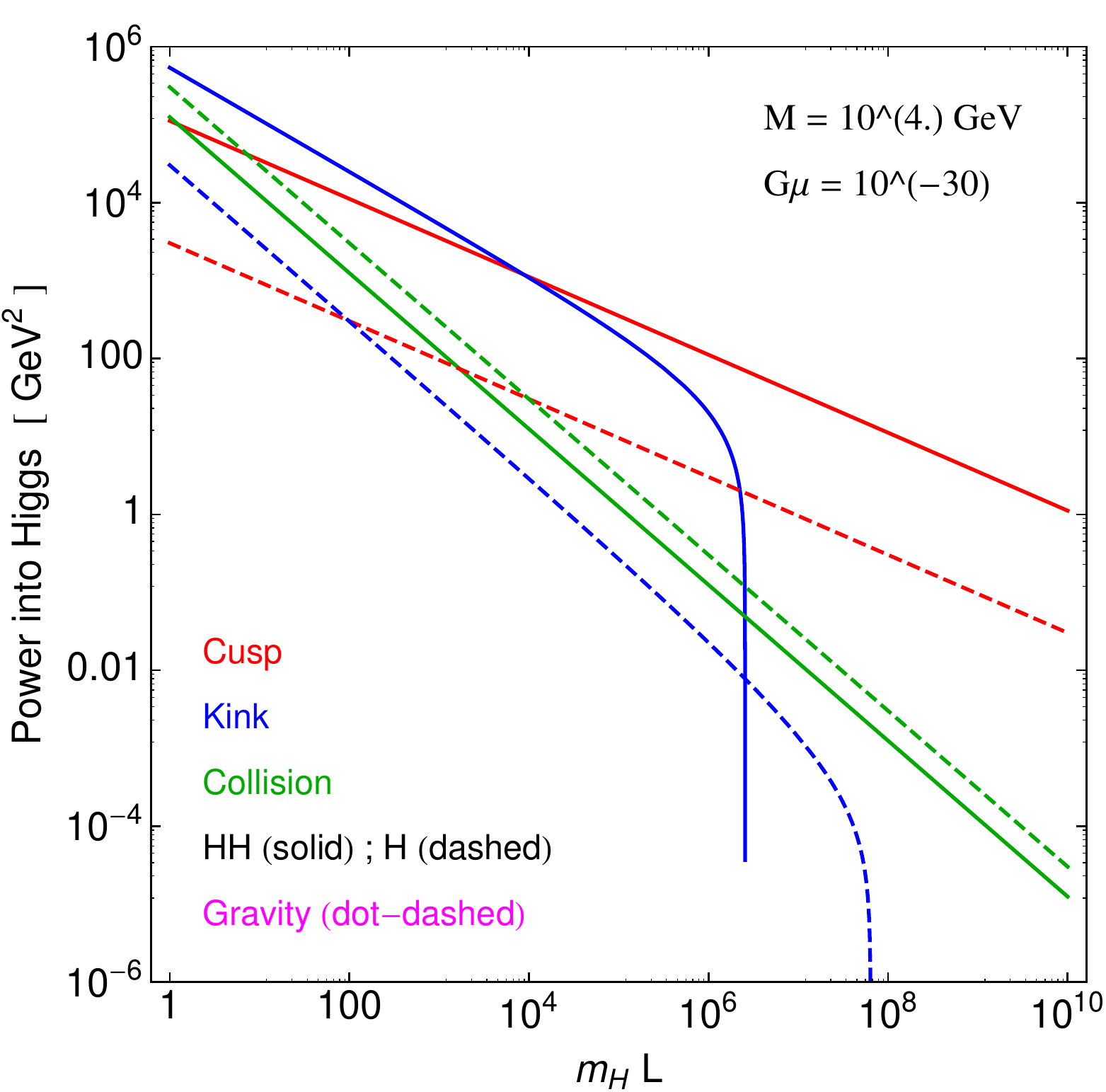} 
\includegraphics[width=0.33\textwidth]{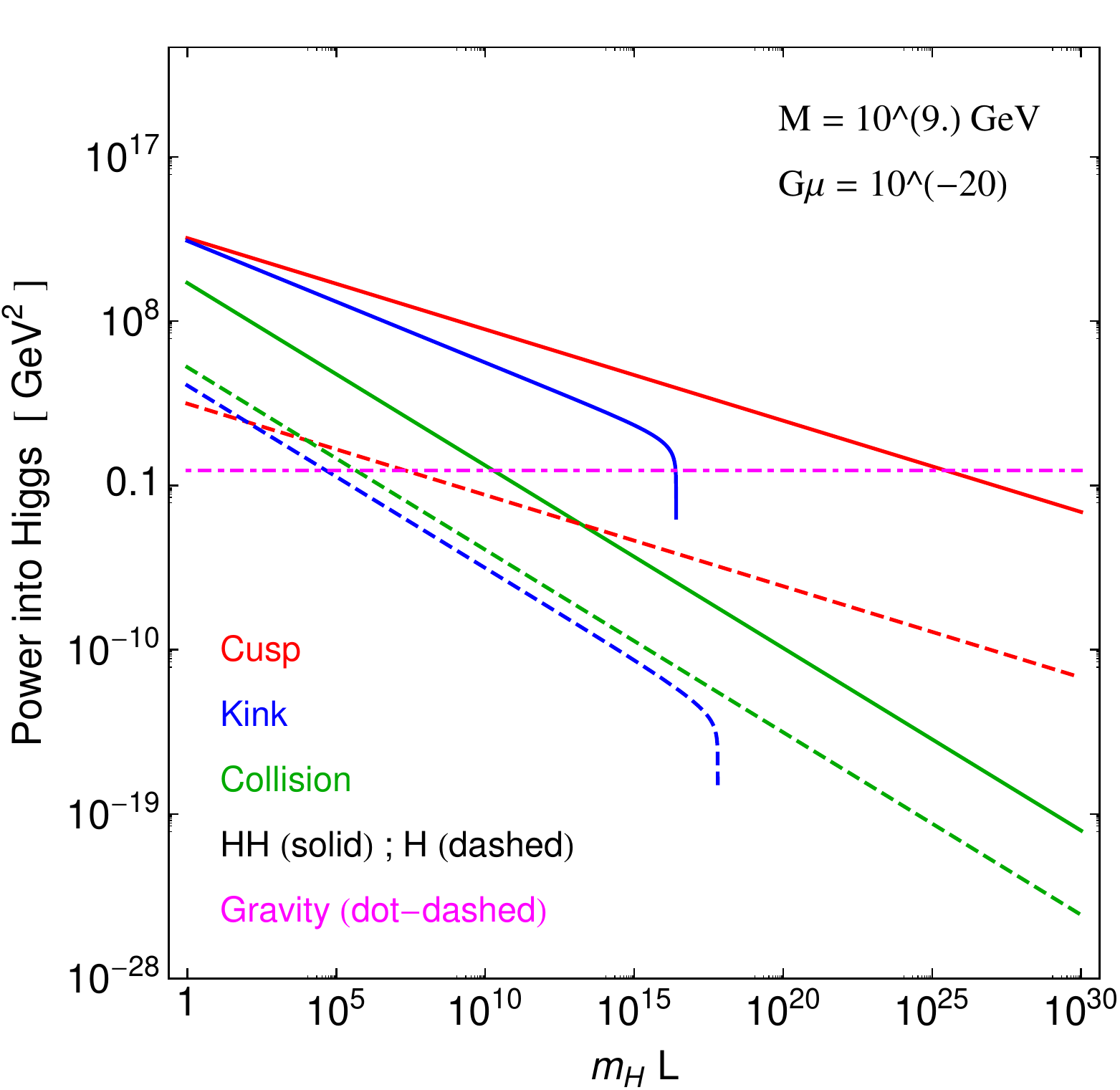} 
\includegraphics[width=0.33\textwidth]{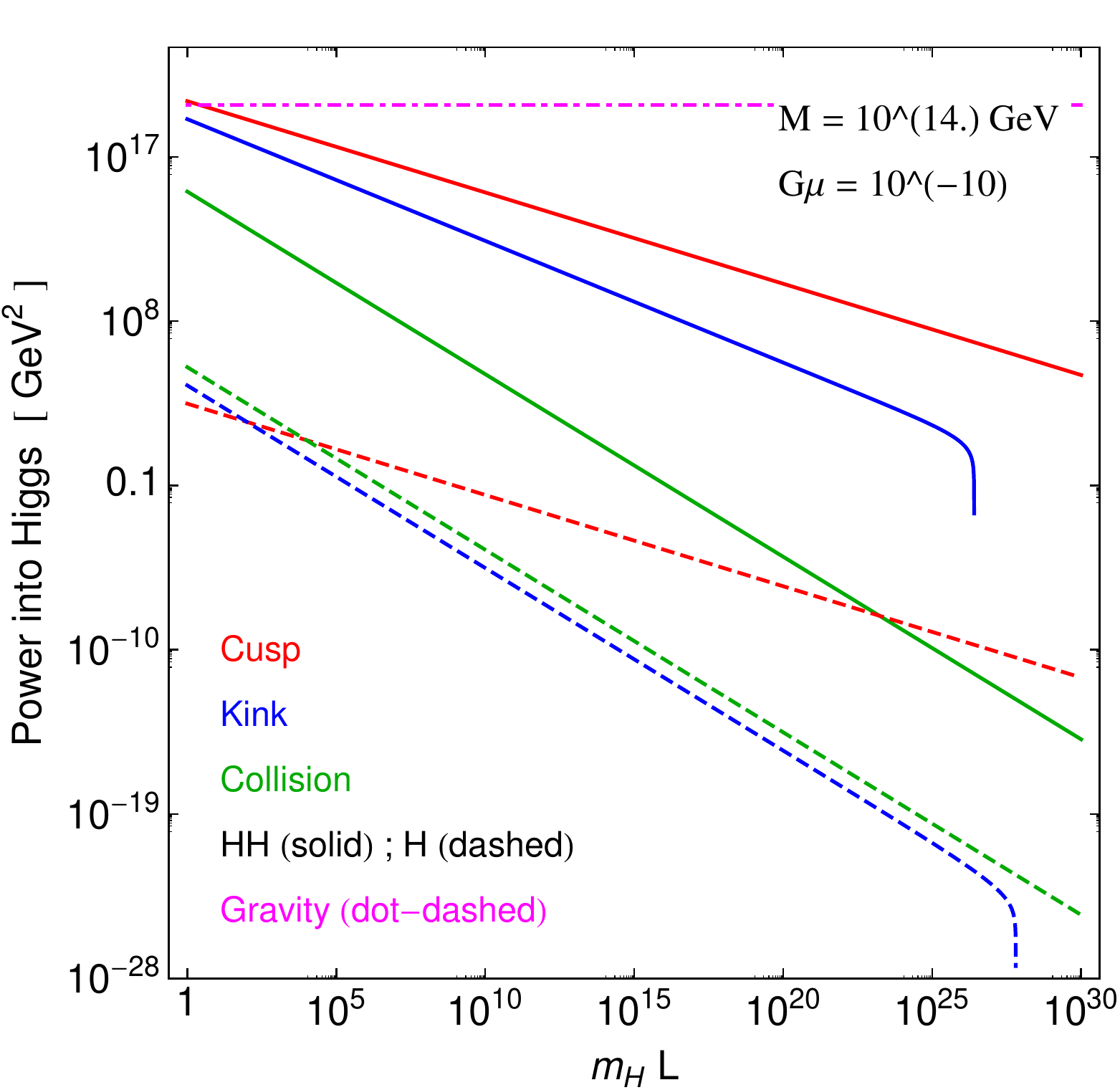} 
\vskip 0.2cm
\caption{
\label{powerplots}
The total power emitted in Higgs radiation from a cusp (red), a kink (blue), and a kink collision (green) due to the linear (dashed) and quadratic (solid) interactions of the Higgs field with the string worldsheet.  
We vary the loop length, $L$, and show three three different string mass scales $M$.  
For reference, $m_H L = 10^{19}$ corresponds to a loop length of $L = 1 \km$, and $L = 40 \, {\rm Gly}$ corresponds to $m_H L = 10^{44}$.   
Note that the scale is different in the left panel.  
}
\end{center}
\end{figure}

Let us highlight the important features of these calculations: 
\begin{enumerate}
	\item  This system is characterized by three hierarchical length scales, the string thickness, the inverse particle mass, and the string loop length: $1 / M \ll 1 / m \ll L$.  The radiation calculation is not amenable to dimensional analysis, because it is always possible to form dimensionless combinations that are far from order one, \eg, $ML \gg 1$ or $m / M \ll 1$.  Additionally, some of the spectra are UV sensitive ($dN = d\kup / \kup^n$ with $n \leq 2$) while others are IR sensitive ($n > 2$), and as a result some of the power formulae depend on the UV mass scale, $M$, while others depend on the IR mass scale, $\eta$, $m_H$, or $m_Z$.  
	\item  In the physically relevant parameter regime, $M L \gg m_H L \gg 1$, the dominant radiation channel is Higgs emission from cuspy loops via the quadratic interaction, see $P_{HH}^{\rm (cusp)}$ in \eref{eq:dom_channel}.  
	\item  The string loop also radiates gravitational waves from cusps, kinks, and kink collisions.  The power output into this channel is well-known: $P_{\rm grav} = \Gamma_{g} G M^4$ where $\mu = M^2$ is the string tension, $\Gamma_{g} \approx 100$, and $G$ is Newton's constant \cite{VilenkinShellard:1994}.  For comparison, $P_{HH}^{\rm (cusp)} \sim M^{3/2} / L^{1/2}$.  If the string mass scale is large, then string loops will primarily radiate in the form of gravitational waves, as originally observed by \rref{Srednicki:1986xg}.  However, it is important to emphasize that particle emission will dominate if the scale of symmetry breaking is low, \eg, for a TeV scale string.  For instance, taking $L \approx 40 \, {\rm Gly}$ to be the size of the horizon today we find $P_{HH}^{\rm cusp} / P_{\rm grav} \approx 10^{4} (M / \TeV)^{-5/2}$.  Moreover, in general Higgs emission dominates over gravitational emission for small loops:  $L < (\GHHcusp)^2 (\gHHstr)^4/ (\Gamma_{g}^2 G^2 M^5)$.  	
	\item  Comparing Higgs emission from a cuspy loop via the linear and quadratic interactions, we find $P_{HH}^{\rm (cusp)} / P_{H}^{\rm (cusp)} \approx (\GHHcusp / \GHcusp) (\gHHstr / \gHstr)^2 (M / m_H)^{3/2}$ where we have approximated $\eta \approx m_H$.  Typically $(\GHHcusp / \GHcusp) \approx 10^{-1}$ and $(\gHHstr / \gHstr) \approx 1$ and $(M / m_H) \gg 1$, and we find that the quadratic interaction is a much more efficient radiation channel.  Note that the dimensionless coefficients (the $\Gamma$ factors) for the quadratic interactions are typically smaller than the corresponding coefficient for the linear interaction; this is a result of the additional phase space suppression (factors of $2\pi$).  
	\item  In \fref{powerplots} we show the six Higgs boson radiation channels.  We use Eqs.~(\ref{eq:IV1a})~to~(\ref{eq:IV2c}) taking $\gHstr = \gHHstr = 1$ and choosing the largest allowed values for the dimensionless prefactors.  In the first panel, the line representing gravitational emission is off the scale of the plot at approximately $10^{-20}$.  For the largest loops, $m_H L \gg 1$, gravitational emission dominates (pink, dot-dashed).  For the smallest loops, $m_H L \approx 1$, the dominant radiation channel is either pair emission from a cusp (red, solid) or pair emission from a kink (blue, solid).  There is no radiation from kinks on large loops, $L \gtrsim M^2 / m_H^3$, since the spectrum is bounded as $m_H \sqrt{m_H L} < \kup < M$.  
	\item  The spectrum of radiation from kinks extends over the range $m \sqrt{mL} < \kup < M$ where $m$ is the particle mass and $M$ is the string mass scale.  For momenta below the IR cutoff, a destructive interference from different segments of the string loop leads to a suppression of radiation.  (In the language of \aref{sub:SaddlePoint}, the saddle point approximation fails.)  For momenta above the UV cutoff, the Compton wavelength of the radiated particle is smaller than the string thickness, $1 / M$, and the radiation is once again suppressed.  (By comparison, the UV cutoff at a cusp is raised to $M \sqrt{ML}$ due to the large boost factor.)  Thus only kinks on small loops, $L < M^2 / m^3$, give appreciable radiation.  
	\item  The Z boson radiation channels are suppressed compared to the corresponding Higgs radiation channels by the fourth or eight power of $(\eta / \sigma) \ll 1$, and this makes Z boson emission negligible.  The factor of $(\eta / \sigma)$ entered the calculation directly in the coupling of the Z boson field to the string, see \eref{eq:Sint}.  For the dark string, the Z boson radiation is only possible by virtue of the gauge-kinetic mixing, and the mixing angle vanishes in the decoupling limit where $(\eta / \sigma) \ll 1$ \cite{Hyde:2013fia}.  For a different model in which this coupling is unsuppressed, the vector boson radiation will be comparable to the Higgs boson radiation, compare \erefs{eq:dom_channel}{PZZcusp_summary}.  
\end{enumerate}

Throughout this analysis we have assumed that the light SM fields are coupled to the zero thickness dark string core, which is composed of the heavy HS fields.  
As we found in \rref{Hyde:2013fia}, the dark string has a much richer structure:  the thin core is surrounded by a wide dressing made up of the SM Higgs and Z boson fields.  
The presence of this dressing could lead to a backreaction that was neglected in our particle production calculations, and this deserves further investigation.  
Additionally, as with most calculations of radiation from cosmic strings, we neglect the more familiar backreaction effect:  a reduction in radiation power as cusps and kinks are gradually smoothed as a result of energy loss in the form of particle and gravitational radiation \cite{Quashnock:1990qy, Quashnock:1990wv}.  

The particle production calculations that we have presented here play a central role in the study of astrophysical and cosmological signatures of cosmic strings.  
For instance, Higgs bosons emitted from the string at late times will decay and produce cosmic rays that are potentially observable on Earth \cite{Vachaspati:2009kq}.  
In our followup paper \cite{Longetal:2014}, we will study the cosmological evolution of the network of dark strings and assess the prospects for their detection.

\acknowledgments
We are very grateful to Yang Bai, Daniel Chung, Dani\`ele Steer, and especially Eray Sabancilar for discussions.  
This work was supported by the Department of Energy at ASU.  

\begin{appendix}

\section{Worldsheet Formalism}\label{app:Worldsheet}

In this appendix we review the string worldsheet formalism (see, \eg, \cite{VilenkinShellard:1994}).  
Let $\tau$ and $\sigma$ be the time-like and space-like worldsheet coordinates, and let $\Xbb^{\mu}(\tau,\sigma)$ be the string worldsheet.  
Then $d^2 \sigma = d\tau d\sigma$ is the worldsheet volume element and $d\sigma^{\mu \nu} = d\tau d\sigma \, \epsilon^{\mu \nu \alpha \beta} \epsilon^{ab} \partial_{a} \Xbb_{\alpha} \partial_{b} \Xbb_{\beta}$ is the worldsheet area element.  
Repeated Greek indices are summed from $0$ to $4$ and Latin indices from $0$ to $1$ with $\partial_{0} \Xbb^{\mu} = \partial_{\tau} \Xbb^{\mu} = \dot{\Xbb}^{\mu}$ and $\partial_{1} \Xbb^{\mu} = \partial_{\sigma} \Xbb^{\mu} = \Xbb^{\mu \, \prime}$.  
We define the pullback of the metric as $\gamma_{ab} \equiv g_{\mu \nu} \partial_{a} \Xbb^{\mu} \partial_{b} \Xbb^{\nu}$ and $\sqrt{-\gamma} \equiv \sqrt{- \det \gamma} = \sqrt{- (1/2) \epsilon^{ab} \epsilon^{cd} \gamma_{ab} \gamma_{cd} }$.  

We now specify to the conformal gauge by imposing
\begin{align}
	\dot{\Xbb} \cdot \Xbb^{\prime} = 0 
	\qquad {\rm and} \qquad
	\dot{\Xbb} \cdot \dot{\Xbb} + \Xbb^{\prime} \cdot \Xbb^{\prime} = 0 \per
\end{align}
Then we have 
\begin{align}
	\label{eq:sigmunu}
	d\sigma^{\mu \nu} 
	& = d \tau d \sigma \, \epsilon^{\mu \nu \alpha \beta} \left( \dot{\Xbb}_{\alpha} \Xbb^{\prime}_{\beta} - \dot{\Xbb}_{\beta} \Xbb^{\prime}_{\alpha} \right) \\
	\label{eq:sqrtgamma}	
	\sqrt{ - \gamma } & =  \big| \dot{\Xbb} \cdot \dot{\Xbb} \bigr| = \dot{\Xbb} \cdot \dot{\Xbb} 
\end{align}
where in the last equality we have used the fact that $\dot{\Xbb}^{\mu}$ is not spacelike.  
We are also free to choose $\tau = t$.  

Solutions of the equation of motion for a free string, $\ddot{\Xbb} = \Xbb^{\prime \prime}$, can be written as 
\begin{align}\label{eq:X_to_ab}
	\Xbb^{\mu}(t,\sigma) &= \frac{1}{2} \bigl[ a^{\mu}(\sigma_{-}) + b^{\mu}(\sigma_{+}) \bigr] 
\end{align}
where we have introduced the right- and left-movers, $a^{\mu}(\sigma_-)$ and $b^{\mu}(\sigma_+)$, which are functions of $\sigma_{\pm} \equiv (\sigma \pm t)$.  
For regularly oscillating string loops, these functions obey the periodicity conditions
\begin{align}\label{eq:ab_periodic}
	a^{\mu}(L + \sigma_{-}) = a^{\mu}(\sigma_{-})
	\qquad {\rm and} \qquad
	b^{\mu}(L + \sigma_{+}) = b^{\mu}(\sigma_{+}) 
\end{align}
in the center of mass frame of the loop.  
The derivatives are 
\begin{align}\label{eq:X_to_ab_derivatives}
	\dot{\Xbb}(t,\sigma) &= \frac{1}{2} \bigl[ - a^{\prime}(\sigma_{-}) + b^{\prime}(\sigma_{+}) \bigr] \nn
	\Xbb^{\prime}(t,\sigma) &= \frac{1}{2} \bigl[ a^{\prime}(\sigma_{-}) + b^{\prime}(\sigma_{+}) \bigr] \\
	\ddot{\Xbb} = \Xbb^{\prime \prime} &= \frac{1}{2} \bigl[ a^{\prime \prime}(\sigma_{-}) + b^{\prime \prime}(\sigma_{+}) \bigr]  \per \nonumber
\end{align}
We can use the residual gauge freedom to choose
\begin{align}\label{eq:ab_param}
\begin{array}{lcl}
	(a)^{\mu} = \bigl\{ -\sigma_{-} \, , \, {\bf a}(\sigma_{-}) \bigr\}
	& \quad , \quad & 
	(b)^{\mu} = \bigl\{ \sigma_{+} \, , \,  {\bf b}(\sigma_{+}) \bigr\} \\
	(a^{\prime})^{\mu} = \bigl\{ -1 \, , \, {\bf a}^{\prime}(\sigma_{-}) \bigr\}
	& \quad , \quad & 
	(b^{\prime})^{\mu} = \bigl\{ 1 \, , \,  {\bf b}^{\prime}(\sigma_{+}) \bigr\} \\
	(a^{\prime \prime})^{\mu} = \bigl\{ 0 \, , \,  {\bf a}^{\prime \prime}(\sigma_{-}) \bigr\}
	& \quad , \quad & 
	(b^{\prime \prime})^{\mu} = \bigl\{ 0 \, , \,  {\bf b}^{\prime \prime}(\sigma_{+}) \bigr\}
\end{array} \com 
\end{align}
along with the condition that $a^{\prime}$ and $b^{\prime}$ should be null, which implies 
\begin{align}\label{eq:12prime_iden}
	&|{\bf a}^{\prime}(\sigma_{-})|^2 = |{\bf b}^{\prime}(\sigma_{+})|^2 = 1 \nn
	&{\bf a}^{\prime} \cdot {\bf a}^{\prime \prime} = {\bf b}^{\prime} \cdot {\bf b}^{\prime \prime} = a^{\prime} \cdot a^{\prime \prime} = b^{\prime} \cdot b^{\prime \prime} = 0 \\
	&{\bf a}^{\prime} \cdot {\bf a}^{\prime \prime \prime} + {\bf a}^{\prime \prime} \cdot {\bf a}^{\prime \prime} 
	= {\bf b}^{\prime} \cdot {\bf b}^{\prime \prime \prime} + {\bf b}^{\prime \prime} \cdot {\bf b}^{\prime \prime} = 0 \per \nonumber
\end{align}
This parametrization lets us write 
\begin{align}\label{eq:measure_identities}
	& d^2 \sigma = d\tau d\sigma = \frac{1}{2} d \sigma_{+} d \sigma_{-} \nn
	& \sqrt{- \gamma} = - \frac{ a^{\prime} \cdot b^{\prime} }{2} \\
	& d\sigma^{\mu \nu} = d\tau d\sigma \, \epsilon^{\mu \nu \alpha \beta} b^{\prime}_{\alpha} a^{\prime}_{\beta} \per \nonumber
\end{align}
Note that $a^{\prime} \cdot b^{\prime} = - 1 - {\bf a}^{\prime} \cdot {\bf b}^{\prime} \leq 0$ and therefore $\sqrt{-\gamma} \geq 0$ as it should be.  

\section{Calculation of Particle Radiation from the String}\label{app:Radiation}

In this appendix, we calculate the spectrum of scalar and vector boson emission due to a coupling with a cosmic string of the linear or quadratic form.  
We also derive the spectrum of fermions emitted due to a direct coupling and an Aharonov-Bohm coupling.  
The results we obtain are not unique to the dark string model; they apply to any model that has couplings of the form considered here.  

We use the matrix element formalism to perform these calculations \cite{Srednicki:1986xg}.  
Since the linear coupling gives rise to a classical source for the scalar or vector field, the radiation in these cases can also be calculated by solving the classical field equation \cite{Vachaspati:2009kq, Sabancilar:2009sq}.  
We have verified that both approaches give identical spectra.  
We also retain all factors of $2$ and $\pi$, which were neglected in the previous calculations.  

\subsection{Scalar Radiation via Linear Coupling}\label{sub:ScalarRad_Linear}

Consider a real scalar field $\phi(x)$ of mass $m$ that is coupled to the string worldsheet $\Xbb^{\mu}(\tau,\sigma)$ through the effective interaction
\begin{align}\label{eq:Hsingle_Leff}
	\Lcal_{\rm eff} & = A \, \phi(x) \, \int d^2 \sigma \sqrt{-\gamma} \, \delta^{(4)}(x - \Xbb) 
\end{align}
where $A$ is an arbitrary real parameter with mass dimension one.  
We calculate the amplitude for particle production by making a perturbative expansion in $A$.  
Then to leading order we have\footnote{More accurately, the initial state is not vacuum, but it is a state containing the string, $\ket{S}$, and the final state contains a deformation of the initial string state, $\ket{S^{\prime}}$.  Provided that the radiation has a negligible backreaction on the string state, one can neglect the deformation and then $\ampl{S^{\prime}}{S} \approx \ampl{S}{S} = \ampl{0}{0}$ \cite{Srednicki:1986xg}.  } 
\begin{align}\label{eq:Hsingle_A}
	\mathcal{A} = i \int d^4 x \, \expval{{\bf k}}{\Lcal_{\rm eff}(x)}{0}
\end{align}
where $\ket{ {\bf k} }$ is a one-particle state of momentum ${\bf k}$.  
The action of the field operator on the one-particle state is simply 
\begin{align}\label{eq:Hsingle_op_on_state}
	\phi(x) \ket{ {\bf k} } = e^{-i k \cdot x} \ket{0}
	\qquad {\rm and} \qquad
	\bra{ {\bf k} } \phi(x) = e^{i k \cdot x} \bra{0}
\end{align}
where $k^{\mu} = \bigl\{ \omega , \, {\bf k} \bigr\}$ with $\omega = \sqrt{ m^2 + \kup^2}$.  
Then upon inserting \eref{eq:Hsingle_Leff} into \eref{eq:Hsingle_A} we obtain 
\begin{align}\label{eq:Hsingle_A_to_I}
	\mathcal{A}(k)
	= i \, A \int d^4 x \, e^{i k \cdot x} \, \int d^2 \sigma \sqrt{-\gamma} \, \delta^{(4)}(x - \Xbb)
	= i \, A \, \mathcal{I}(k)
\end{align}
where
\begin{align}\label{eq:Hsingle_I}
	\mathcal{I}(k) \equiv \int d^2 \sigma \sqrt{-\gamma}  \, e^{i k \cdot \Xbb}  \per 
\end{align}
In \aref{app:ST_integrals} we calculate this integral for various string configurations, as specified by $\Xbb^{\mu}(\tau,\sigma)$.  

For a given $\Xbb$ we calculate the number of scalar bosons emitted into a phase space volume $d^3 k = \kup^2 d\kup \, d \Omega$ as 
\begin{align}\label{eq:Hsingle_dN_def}
	dN
	= \frac{d^3 k}{(2\pi)^3 2 \omega} | \mathcal{A}(k) |^2 \per
\end{align}
Using \eref{eq:Hsingle_A_to_I} in \eref{eq:Hsingle_dN_def} we obtain the final spectrum
\begin{align}\label{eq:Hsingle_dN}
	dN
	= A^2 \, \frac{d^3 k}{(2\pi)^3 2 \omega} | \mathcal{I}(k) |^2 
\end{align}
where the dimensionful coefficient is equal to $A = \gHstr \eta$ for the dark string.

\subsection{Scalar Radiation via Quadratic Coupling}\label{sub:ScalarRad_Quad}

Consider a real scalar field $\phi(x)$ of mass $m$ that is coupled to the string worldsheet $\Xbb^{\mu}(\tau,\sigma)$ through the effective interaction
\begin{align}\label{eq:Hquad_Leff}
	\Lcal_{\rm eff} & = C \, \phi(x)^2 \, \int d^2 \sigma \, \sqrt{-\gamma} \, \delta^{(4)}(x - \Xbb) 
\end{align}
where $C$ is an arbitrary real parameter with mass dimension zero.  
We can calculate the radiation of scalar boson pairs using perturbation theory provided that $C \ll 1$.  
Consider the radiation of a boson pair with momenta ${\bf k}$ and $\bar{\bf k}$.  
We can introduce the 4-vectors $k^{\mu} = \bigl\{ \omega = \sqrt{ {\bf k}^2 + m^2 } \, , \, {\bf k} \bigr\}$ and $\bar{k}^{\mu} = \bigl\{ \bar{\omega} = \sqrt{ \bar{\bf k}^2 + m^2 } \, , \, \bar{\bf k} \bigr\}$.  
To leading order in $C$ the amplitude for this process is 
\begin{align}\label{eq:Hquad_A}
	\mathcal{A} = i \int d^4 x \, \expval{{\bf k} \, \bar{\bf k}}{\Lcal_{\rm eff}(x)}{0} \per
\end{align}
Inserting \eref{eq:Hquad_Leff} into \eref{eq:Hquad_A} and using \eref{eq:Hsingle_op_on_state} we obtain 
\begin{align}\label{eq:Hquad_A_to_I}
	\mathcal{A} & 
	= i C \, \mathcal{I}(k+\bar{k})
\end{align}
where $\Ical(k)$ was defined in \eref{eq:Hsingle_I}.  
The number of scalar bosons emitted into the phase space volume $d^3k \, d^3 \bar{k} = \kup^2 \, d\kup \, d\Omega \, \bar{\kup}^2 \, d\bar{\kup} \, d\bar{\Omega}$ is calculated as 
\begin{align}\label{eq:Hquad_dN_def}
	dN = \frac{d^3 k}{(2\pi)^3 2 \omega} \frac{d^3 \bar{k}}{(2\pi)^3 2 \bar{\omega}} \abs{\mathcal{A}}^2 \per
\end{align}
Using \eref{eq:Hquad_A_to_I} this becomes
\begin{align}\label{eq:Hquad_dN}
	dN = C^2 \, \frac{d^3 k}{(2\pi)^3 2 \omega} \frac{d^3 \bar{k}}{(2\pi)^3 2 \bar{\omega}} \abs{\mathcal{I}(k+\bar{k})}^2 
\end{align}
where $C =\gHHstr$ for the dark string.

\subsection{Vector Radiation via Linear Coupling}\label{sub:VectorRad_Linear}

Consider a vector field $A_{\mu}(x)$ of mass $m$ that couples to the string worldsheet $\Xbb^{\mu}(\tau,\sigma)$, via the linear interaction 
\begin{align}\label{eq:Zsingle_Leff}
	\Lcal_{\rm eff} = \frac{C}{2} F_{\mu\nu}(x) \int d\sigma^{\mu\nu} \delta^{(4)}(x-\Xbb)
\end{align}
where $F_{\mu \nu} = \partial_{\mu} A_{\nu} - \partial_{\nu} A_{\mu}$ is the field strength tensor and $C$ is a real parameter of mass dimension zero.  
Recall that the worldsheet area element was defined in \eref{eq:sigmunu}.  
Since the radiation will be relativistic, we can treat the gauge boson as transversely polarized with two allowed helicties $\lambda = \pm 1$.  
We calculate the amplitude to radiate a vector boson with momentum ${\bf k}$ and helicity $\lambda$ as 
\begin{align}\label{eq:Zsingle_A}
	\mathcal{A} = i \int d^4x \, \expval{ {\bf k} , \lambda }{ \Lcal_{\rm eff}(x) }{ 0 } \per
\end{align}
The action of the field operator on the one-particle state is 
\begin{align}\label{eq:Zsingle_A_on_state}
	A_{\nu}(x) \ket{{\bf k},\lambda} = \epsilon_{\nu}(k,\lambda) e^{-ik \cdot x} \ket{0}
	\qquad {\rm and} \qquad
	\bra{{\bf k},\lambda} A_{\nu}(x) = \epsilon_{\nu}^{\ast}(k,\lambda) e^{ik \cdot x} \bra{0}
\end{align}
where $k^{\mu} = \bigl\{ \omega = \sqrt{m^2 + \kup^2} \, , \, {\bf k} \bigr\}$.  
Inserting \eref{eq:Zsingle_Leff} into \eref{eq:Zsingle_A} and using \eref{eq:Zsingle_A_on_state} gives
\begin{align}\label{eq:Zsingle_A_to_I}
	\mathcal{A} 
	= - C \, k_{\mu} \epsilon^{\ast}_{\nu}(k,\lambda) \, \mathcal{I}^{\mu \nu}(k)
\end{align}
where
\begin{align}\label{eq:Zsingle_I}
	\mathcal{I}^{\mu \nu}(k) \equiv \int d\sigma^{\mu \nu} \, e^{i k \cdot \Xbb} \per
\end{align}
Then the number of vector bosons emitted into the phase space volume $d^3 k = \kup^2 \, d\kup \, d\Omega$ is calculated as 
\begin{align}\label{eq:Zsingle_dN_def}
	dN 
	= \sum_{\lambda} \frac{d^3k}{(2\pi)^3 2\omega} \, | \mathcal{A} |^2 
\end{align}
where we sum over the two polarization states.  
Using \eref{eq:Zsingle_A_to_I} this becomes
\begin{align}\label{eq:Zsingle_dN_before_completeness}
	dN = C^2 \frac{d^3k}{(2\pi)^3 2\omega} \, k_{\mu} k_{\alpha} \ \mathcal{I}^{\mu \nu}(k) \ \mathcal{I}^{\alpha \beta}(k)^{\ast} \ \sum_{\lambda} \epsilon_{\beta}(k,\lambda) \epsilon_{\nu}^{\ast}(k,\lambda) \per
\end{align}
We perform the spin sum using the completeness relationship 
\begin{align}\label{eq:Zsingle_completeness}
	\sum_{\lambda=\pm1} \epsilon_{\beta}(k,\lambda) \epsilon_{\nu}^{\ast}(k,\lambda) = - g_{\beta \nu} 
	\per  
\end{align}
Doing so we find the spectrum to be 
\begin{align}\label{eq:Zsingle_dN}
	dN = C^2 \frac{d^3k}{(2\pi)^3 2\omega} \, k^2 \, \Pi(k)
\end{align}
where the (positive, real) function
\begin{align}\label{eq:Pi_def}
	\Pi(q) \equiv - g_{\nu\beta} \frac{q_{\mu} q_{\alpha}}{q^2} \ \mathcal{I}^{\mu \nu}(q) \mathcal{I}^{\alpha \beta}(q)^{\ast} 
\end{align}
has dimensions of $\text{length}^4$ and carries the dependence on the string worldsheet.  
By choosing $C = \gZstr ( \eta / \sigma )^2$ we obtain the spectrum of Z boson radiation from the dark string.

\subsection{Vector Radiation via Quadratic Coupling}\label{sub:VectorRad_Quad}

Consider a vector field $A^{\mu}$ of mass $m$ that couples to the string worldsheet via the quadratic interaction
\begin{align}\label{eq:Zquad_Leff}
	\Lcal_{\rm eff} = C \, A_{\mu}(x) A^{\mu}(x) \int d^2 \sigma \sqrt{-\gamma} \, \delta^{(4)}(x-\Xbb)
\end{align}
where $C$ is a real parameter of mass dimension zero.  
The amplitude to radiate a pair of vector bosons with momenta ${\bf k}$ and $\bar{\bf k}$ and helicities $\lambda$ and $\bar{\lambda}$ is calculated as 
\begin{align}\label{eq:Zquad_A}
	\mathcal{A} = i \int d^4x \, \expval{ {\bf k} , \lambda ; \bar{\bf k} , \bar{\lambda} }{ \Lcal_{\rm eff}(x) }{ 0 } \per
\end{align}
We can introduce the 4-vectors $k^{\mu} = \bigl\{ \omega = \sqrt{ {\bf k}^2 + m^2 } \, , \, {\bf k} \bigr\}$ and $\bar{k}^{\mu} = \bigl\{ \bar{\omega} = \sqrt{ \bar{\bf k}^2 + m^2 } \, , \, \bar{\bf k} \bigr\}$.  
Upon inserting \eref{eq:Zquad_Leff} into \eref{eq:Zquad_A} and using \eref{eq:Zsingle_A_on_state} we obtain
\begin{align}\label{eq:Zquad_A_to_I}
	\mathcal{A} 
	& = i C \, \epsilon^{\ast}_{\mu}(k,s) \, \epsilon^{\ast}_{\nu}(\bar{k},\bar{s}) \, g^{\mu \nu} \, \mathcal{I}(k + \bar{k})
\end{align}
where $\Ical(k)$ was defined in \eref{eq:Hsingle_I}.  
Then the number of vector bosons emitted into the phase space volume $d^3 k \, d^3 \bar{k} = \kup^2 \, d\kup \, d\Omega \, \bar{\kup}^2 \, d\bar{\kup} \, d\bar{\Omega}$ is calculated as 
\begin{align}\label{eq:Zquad_dN_def}
	dN 
	= \sum_{\lambda} \sum_{\bar{\lambda}} \frac{d^3k}{(2\pi)^3 2\omega} \, \frac{d^3\bar{k}}{(2\pi)^3 2\bar{\omega}} \, | \mathcal{A} |^2 
\end{align}
where we sum over the transverse polarization states $\lambda , \bar{\lambda} = \pm 1$.  
(Since the radiation is highly boosted, we can neglect the longitudinal polarization states.)  
Using \eref{eq:Zquad_A_to_I} this becomes
\begin{align}\label{eq:Zquad_dN_before_completeness}
	dN = C^2 \frac{d^3k}{(2\pi)^3 2\omega} \, \frac{d^3\bar{k}}{(2\pi)^3 2\bar{\omega}} \, | \mathcal{I}(k + \bar{k}) |^2 \ g^{\mu \nu} g^{\alpha \beta} \sum_{\lambda} \epsilon_{\alpha}(k,\lambda) \epsilon_{\mu}^{\ast}(k,\lambda) \ \sum_{\lambda} \epsilon_{\beta}(\bar{k},\bar{\lambda}) \epsilon_{\nu}^{\ast}(\bar{k},\bar{\lambda}) \per
\end{align}
We evaluate the spin sums using the completeness relation in \eref{eq:Zsingle_completeness} to find 
\begin{align}\label{eq:Zquad_dN}
	dN 
	& = 4 \, C^2 \frac{d^3k}{(2\pi)^3 2\omega} \, \frac{d^3\bar{k}}{(2\pi)^3 2\bar{\omega}} \, | \mathcal{I}(k + \bar{k}) |^2 
	\per
\end{align}
For the dark string model we take $C = \gZZstr (\eta / \sigma)^4$.  

\subsection{Dirac Spinor Radiation -- Direct Coupling}\label{sub:SpinorRad_1}

Consider a Dirac field $\Psi(x)$ of mass $m$ that is coupled to the string worldsheet $\Xbb^{\mu}(\tau,\sigma)$ through the effective interaction
\begin{align}\label{eq:Psi_Leff}
	\Lcal_{\rm eff} & = \frac{C}{M} \, \bar{\Psi}(x) \Psi(x) \, \int d^2 \sigma \, \sqrt{-\gamma} \, \delta^{(4)}(x - \Xbb) 
\end{align}
where $C$ is an arbitrary real parameter with mass dimension zero, and $M$ is the string mass scale.  
Consider the radiation of a particle / anti-particle pair with momenta ${\bf k}$ and $\bar{\bf k}$ and spins $s$ and $\bar{s}$.  
We can introduce the 4-vectors $k^{\mu} = \bigl\{ \omega = \sqrt{ {\bf k}^2 + m^2 } \, , \, {\bf k} \bigr\}$ and $\bar{k}^{\mu} = \bigl\{ \bar{\omega} = \sqrt{ \bar{\bf k}^2 + m^2 } \, , \, \bar{\bf k} \bigr\}$.  
To leading order the amplitude for this process is 
\begin{align}\label{eq:Psi_A}
	\mathcal{A} = i \int d^4 x \, \expval{{\bf k} , s \, ; \, \bar{\bf k} , \bar{s} }{\Lcal_{\rm eff}(x)}{0} \per
\end{align}
The action of the field operator on the one-particle state is given by
\begin{align}
	\bra{{\bf k}, s} \bar{\Psi}(x) = \bar{u}({\bf k},s) e^{i k \cdot x} \bra{0}
	\qquad {\rm and} \qquad 
	\bra{\bar{\bf k}, \bar{s}} \Psi(x) = v(\bar{\bf k}, \bar{s}) e^{i \bar{k} \cdot x} \bra{0} \per
\end{align}
Inserting \eref{eq:Psi_Leff} into \eref{eq:Psi_A} we obtain 
\begin{align}\label{eq:Psi_A_to_I}
	\mathcal{A} & = i \frac{C}{M} \bar{u}({\bf k},s) v(\bar{\bf k}, \bar{s}) \mathcal{I}(k + \bar{k})
\end{align}
where $\Ical(k)$ was defined in \eref{eq:Hsingle_I}.  
The number of particle pairs emitted into the phase space volume $d^3k \, d^3 \bar{k} = \kup^2 \, d\kup \, d\Omega \, \bar{\kup}^2 \, d\bar{\kup} \, d\bar{\Omega}$ is calculated as in \eref{eq:Zquad_dN_def} where now the sum is over spin states $s, \bar{s} = \pm 1/2$.  
We use the completeness relations, 
\begin{align}
	\sum_{s} u({\bf k},s) \bar{u}({\bf k},s) = (k_{\mu} \gamma^{\mu} + m)
	\qquad {\rm and} \qquad
	\sum_{\bar{s}} v(\bar{\bf k}, \bar{s}) \bar{v}(\bar{\bf k}, \bar{s}) = (\bar{k}_{\mu} \gamma^{\mu} - m) \per
\end{align}
Using the familiar Dirac gamma trace relations, we obtain 
\begin{align}\label{eq:Psi_dN}
	dN = 4C^2 \, \frac{d^3 k}{(2\pi)^3 2 \omega} \frac{d^3 \bar{k}}{(2\pi)^3 2 \bar{\omega}} \abs{\mathcal{I}(k+\bar{k})}^2 \frac{k \cdot \bar{k} - m^2}{M^2}
\end{align}
where $C =\gpsistr (\eta/\sigma)^2$ for the dark string.  

\subsection{Dirac Spinor Radiation -- AB Coupling}\label{sub:SpinorRad_2}

Consider a Dirac field $\Psi(x)$ of mass $m$ that is coupled to the string worldsheet $\Xbb^{\mu}(\tau,\sigma)$ through the effective interaction
\begin{align}\label{eq:Psi_Leff_2}
	\Lcal_{\rm eff} & = C \, \bar{\Psi}(x) \gamma_{\mu} \Psi(x) \int_{\rm ret.} \frac{d^4 p}{(2\pi)^4} \frac{i p_{\nu}}{p^2} \Ical^{\mu \nu}(p) e^{-i p \cdot x} 
\end{align}
where $C \ll 1$ is an arbitrary real parameter with mass dimension zero, and $\Ical^{\mu \nu}(k)$ was defined in \eref{eq:Zsingle_I}.  
In the momentum integral, the integration contour is extended above both poles at $p^0 = \pm \abs{\bf p}$.  
Following \sref{sub:SpinorRad_1} we calculate the amplitude for the radiation of a particle/anti-particle pair:  
\begin{align}\label{eq:Psi_A_to_I_2}
	\mathcal{A} & = - C \, \bar{u}({\bf k},s) \gamma_{\mu} v(\bar{\bf k}, \bar{s}) \, \frac{q_{\nu}}{q^2} \Ical^{\mu \nu}(q) 
\end{align}
where $q \equiv k + \bar{k}$.  
The number of particle pairs emitted into the phase space volume $d^3k \, d^3 \bar{k} = \kup^2 \, d\kup \, d\Omega \, \bar{\kup}^2 \, d\bar{\kup} \, d\bar{\Omega}$ is calculated as in \eref{eq:Zquad_dN_def} where now the sum is over spin states $s, \bar{s} = \pm 1/2$. 
Using \eref{eq:Psi_A_to_I_2} we obtain 
\begin{align}
	dN & = 4 C^2 \, \frac{d^3 k}{(2\pi)^3 2 \omega} \frac{d^3 \bar{k}}{(2\pi)^3 2 \bar{\omega}} \left( k_{\mu} \bar{k}_{\alpha} + k_{\alpha} \bar{k}_{\mu} - (m^2 + k \cdot \bar{k}) g_{\mu \alpha} \right) \frac{ q_{\nu} q_{\beta} }{ q^4 } 
	\Ical^{\mu \nu}(q)
	\Ical^{\alpha \beta}(q)^{\ast} \per
\end{align}
Then using the antisymmetry of $\Ical^{\mu \nu}$ we find 
\begin{align}
\label{dNAB}
	dN & = 2 C^2 \, \frac{d^3 k}{(2\pi)^3 2 \omega} \frac{d^3 \bar{k}}{(2\pi)^3 2 \bar{\omega}} 
	\Bigl[
	\Pi(k + \bar{k}) - | \Upsilon(k,\bar{k}) |^2
	\Bigr]
\end{align}
where $\Pi(q)$ is defined in \eref{eq:Pi_def} and 
\begin{align}\label{eq:Upsilon_def}
	\Upsilon(k,\bar{k}) \equiv \frac{ 2 k_{\mu} \bar{k}_{\nu} }{ ( k + \bar{k})^2 } \Ical^{\mu \nu}(k + \bar{k}) \per 
\end{align}
In general, the evaluation of \eref{dNAB} is very involved and must be done numerically for some
choice of loops as in \cite{Chu:2010zzb}.  
However, to extract the radiation spectrum, it is sufficient to note that $dN > 0$, and so the term containing $\Upsilon$ is never larger than the term containing $\Pi$ [see also \eref{eq:Upsilon_to_Pi}].  
Hence, to extract scalings, we will take\footnote{
There is a danger that there can be cancellations
between the $\Pi$ and $|\Upsilon |^2$ terms but we find that our scalings agree with the behavior that 
was numerically obtained in \cite{Chu:2010zzb} for similar loops.  }
\begin{align}\label{eq:Psi_dN_2}
	dN & \approx 2 C^2 \, \frac{d^3 k}{(2\pi)^3 2 \omega} \frac{d^3 \bar{k}}{(2\pi)^3 2 \bar{\omega}} \Pi(k + \bar{k})
\end{align}
where for the dark string $C = - (2\pi \theta_{q}) / 2$.  

\section{Calculation of the Worldsheet Integrals}\label{app:Worldsheet_Integrals}

In \aref{app:Radiation} we encountered the two integrals 
\begin{align}
	\label{eq:IH_def_appendix} 
	&  \Ical(k) = \int d^2 \sigma \, \sqrt{-\gamma} \, e^{i k \cdot \Xbb} \\
	\label{eq:IZ_def_appendix}
	&  \Ical^{\mu \nu}(k) = \int d^2 \sigma^{\mu \nu} \, e^{i k \cdot \Xbb} 
\end{align}
while calculating the radiation spectra.  
In this appendix and the next, we will analytically calculate these integrals for the cusp, kink, and kink-kink collision string configurations.  

It is convenient to define the integrals 
\begin{align}\label{eq:Ipm_def}
	\Ical^{\mu}_{+}(b; k) \equiv \int_{0}^{L} d \sigma_{+} \, b^{\prime \, \mu} e^{i k \cdot b / 2} 
	\qquad {\rm and} \qquad 
	\Ical^{\mu}_{-}(a; k) \equiv \int_{0}^{L} d \sigma_{-} \, a^{\prime \, \mu} e^{i k \cdot a / 2} 
\end{align}
where $\Ical_{+}$ is a functional of $b^{\mu}(\sigma_+)$ with parameter $k^{\mu}$, and similarly $\Ical_{-}$ is a functional of $a^{\mu}(\sigma_{-})$.  
For a regularly oscillating string loop, the periodicity of $a^{\prime}$ and $b^{\prime}$ implies the identities
\begin{align}\label{eq:Ipm_iden}
	k \cdot \Ical_{\pm} = 0 \per
\end{align}
Additionally, for such a loop we can factorize the original integrals from \erefs{eq:IH_def_appendix}{eq:IZ_def_appendix} in terms of $\Ical_{+}$ and $\Ical_{-}$.  
We use \eref{eq:measure_identities} to factor the integrands, and we use the periodicity of the loop oscillation to rewrite the domain of integration as $\int_{0}^{L} d \sigma \int_{0}^{T} d\tau = (1/2) \int_{0}^{L} d\sigma_{+} \int_{0}^{L} d\sigma_{-}$ where $T = L/2$ is the loop oscillation period.  
Doing so gives 
\begin{align}
	\label{eq:IcalH}
	&  \Ical(k) = - \frac{1}{4} \, g^{\alpha \beta} (\Ical_{+}(b;k))_{\alpha} \, (\Ical_{-}(a;k))_{\beta} \\
	\label{eq:IcalZ}
	&  \Ical^{\mu \nu}(k) = \frac{1}{2} \epsilon^{\mu \nu \alpha \beta} ( \Ical_{+}(b;k))_{\alpha} \, (\Ical_{-}(a;k) )_{\beta} \per
\end{align}
The problem is now reduced to calculating the two integrals, $\Ical_{+}^{\mu}$ and $\Ical_{-}^{\mu}$, for a given loop configuration, specified by $a^{\mu}$ and $b^{\mu}$.  

These integrals cannot be performed analytically for general configurations.  
We, therefore, focus on the configurations that we expect to maximize the integrals, since this corresponds to maximum particle radiation.  
It turns out that for these optimum configurations, the saddle point and the discontinuity, the integrals are analytically tractable.  

\subsection{Saddle Point Integral}\label{sub:SaddlePoint}

The integrals in \eref{eq:Ipm_def} become analytically tractable if there is a saddle point at which the phase is stationary \cite{Damour:2001bk}.  
For the sake of discussion consider the integral $\Ical_{+}$.  
Its phase can be expanded about $\sigma_{+} = \sigma_{s}$ as 
\begin{align}\label{eq:phase}
	\frac{k \cdot b(\sigma_+)}{2} & = \frac{k \cdot b_{s}}{2} + \frac{k \cdot b_{s}^{\prime}}{2} \, (\sigma_+ - \sigma_{s}) + \frac{k \cdot b_{s}^{\prime \prime}}{4} \, (\sigma_+ - \sigma_{s})^2 + \frac{k \cdot b_{s}^{\prime \prime \prime}}{12} \, (\sigma_+ - \sigma_{s})^3 + \ldots \per
\end{align}
Subscripts are used to denote evaluation of the function at a particular point, \eg, $b_{s}^{\prime} = b^{\prime}(\sigma_{s})$.  
We say that $\sigma_{s}$ is a saddle point if the stationary phase criterion, 
\begin{align}\label{eq:sp_condit}
	k \cdot b^{\prime}_{s} = 0 \com
\end{align}
is satisfied.  
Using \eref{eq:ab_param} this can be written as 
\begin{align}
	k \cdot b^{\prime}_{s} = \omega - {\bf k} \cdot {\bf b}^{\prime}_{s} = \omega - \kup \, \cos \theta 
\end{align}
where $\theta$ is the angle between ${\bf k}$ and ${\bf b}^{\prime}_{s}$.  
If the particle being radiated is massless, $\omega = \kup$, then the saddle point criterion is satisfied by choosing ${\bf k} = \kup \, {\bf b}_{s}^{\prime}$ (\ie, $\theta = 0$).  
Then it follows from the identity in \eref{eq:12prime_iden} that $k \cdot b^{\prime \prime}_{s} = 0$ as well, and the leading term in \eref{eq:phase} is cubic.  

For massive particle radiation the saddle point criterion cannot be satisfied exactly.  
Instead, we have instead a quasi-saddle point, $\sigma_{+} = \sigma_{\qsp}$, at which the phase is approximately stationary:  
\begin{align}\label{eq:qsp_condit}
	{\bf k} = \kup \, {\bf b}^{\prime}_{\qsp} 
	\quad , \quad
	k \cdot b_{\qsp}^{\prime} = \omega - \kup 
	\quad , \quad
	k \cdot b^{\prime \prime}_{\qsp} = 0 
	\quad , \quad
	k \cdot b^{\prime \prime \prime}_{\qsp} = \kup \, | {\bf b}_{\qsp}^{\prime \prime} |^2 \com
\end{align}
where we have used \eref{eq:12prime_iden}.  
It will be convenient to write 
\begin{align}\label{eq:shape_param}
	{\bf b}^{\prime \prime}_{\rm qsp} = \frac{2\pi}{L} \beta_{\rm qsp} \hat{\bf b}^{\prime \prime}_{\rm qsp}
	\qquad {\rm and} \qquad
	{\bf a}^{\prime \prime}_{\rm qsp} = \frac{2\pi}{L} \alpha_{\rm qsp} \hat{\bf a}^{\prime \prime}_{\rm qsp}
\end{align}
where the hatted quantities are unit vectors.  
The dimensionless parameters $\alpha_{\rm asp}$ and $\beta_{\rm asp}$ are related to the acceleration or curvature of the loop at the quasi-saddle point (recall \eref{eq:X_to_ab_derivatives}).  
The stationary phase approximation is still applicable as long as $(k \cdot b_{\rm qsp}^{\prime}) (\sigma_{+}) \ll (k \cdot b_{\rm qsp})^{\prime \prime \prime} (\sigma_{+})^3 \ll 2 \pi$.  

Suppose that we are given a configuration $b^{\mu}(\sigma_{+})$ and a $k^{\mu}$ such that there exists some point $\sigma_+ = \sigma_{s}$ where the quasi-saddle point condition, \eref{eq:qsp_condit}, is satisfied.  
Then the integral from \eref{eq:Ipm_def} can be approximated by expanding in $\Delta \sigma = \sigma_{+} - \sigma_{s}$, which gives  
\begin{align}\label{eq:sp_integral}
	\Ical_{+}^{\mu}(b;k) \approx e^{i \frac{k \cdot b_{s}}{2}} \int_{-\sigma_{s}}^{L - \sigma_{s}} d (\Delta \sigma) \, \left[ (b^{\prime}_{s})^{\mu} + (b^{\prime \prime}_{s})^{\mu} \Delta \sigma \right] \exp{i \phi(\Delta \sigma) } \per
\end{align}
The phase is also expanded in powers of $\Delta \sigma / L$ as $\phi(\Delta \sigma) = \phi_{1}(\Delta \sigma) + \phi_{3}(\Delta \sigma) + \ldots $ where 
\begin{align}\label{eq:sp_phase}
	&\phi_1 \equiv \frac{\omega-\kup}{2} \Delta \sigma 
	\qquad {\rm and} \qquad
	\phi_3 \equiv \frac{2\pi \kup}{L^2} \frac{1}{\Theta^3} (\Delta \sigma)^3 \per
\end{align}
Here we have introduced the dimensionless parameter $\Theta \equiv (6 / \pi \beta_s^2)^{1/3}$, and the shape parameter is $\beta_s = L \abs{ {\bf b}_{s}^{\prime \prime} }/(2\pi )$ as per \eref{eq:shape_param}.  

As long as $\phi_1$ is negligible, the integral is in the stationary phase regime, and it can be evaluated directly with the saddle point approximation.  
Since the integral is dominated by the saddle point, we can extend the limits of integration to infinity.  
Doing so we obtain 
\begin{align}
	\Ical_{+}(b;k) & \approx 
	e^{i \frac{k \cdot b_{s}}{2}}  
	\int_{-\infty}^{\infty} d(\Delta \sigma) \left[ b^{\prime}_{s} + b^{\prime \prime}_{s} \Delta \sigma \right] \exp{i \phi_{3}(\Delta \sigma) }   \nn
	& = e^{i \frac{k \cdot b_{s}}{2}}  L \left( 
	A_{+} \frac{b^{\prime}_{s}}{(\kup L)^{1/3}} 
	+ i B_{+} \frac{L b^{\prime \prime}_{s}}{(\kup L)^{2/3}} 
	\right)
	\label{eq:Iplus_sp_final}
\end{align}
where
\begin{align}\label{eq:AB_def}
	A_{+} = \frac{(2\pi)^{2/3}}{3\Gamma(2/3)} \Theta
	\qquad {\rm and} \qquad
	B_{+} = \frac{\Gamma(2/3)}{\sqrt{3}} \frac{\Theta^2}{(2\pi)^{2/3}} \com
\end{align}
and $\Theta = (6 / \pi \beta_s^2)^{1/3}$ was defined in the paragraph above.  

The linear phase, $\phi_1$, must be negligible if the saddle point approximation is to be valid.  
We define the ``width of the saddle point'' by the condition $\phi_3(\Delta \sigma_{\rm max}) = 2\pi$, which gives
\begin{align}\label{eq:Dsigma_max}
	\Delta \sigma_{\rm max} = \Theta L (\kup L)^{-1/3} \per
\end{align}
Imposing $\phi_1(\Delta \sigma_{\rm max}) < \phi_{3}(\Delta \sigma_{\rm max})$ leads to the bound 
\begin{align}\label{eq:general_bound}
	\frac{\Theta}{4\pi} L^{2/3} (\omega - \kup) < \kup^{1/3} \per
\end{align}
The left-hand side vanishes in the relativistic limit, and the bound becomes saturated as the momentum is lowered.  
Approximating $\omega \approx \kup + m^2 / 2\kup$ we obtain a lower bound on the momentum \cite{Srednicki:1986xg}
\begin{align}\label{eq:kmin_bound}
	\kap \, m \sqrt{mL} < \kup
	\qquad {\rm with} \qquad
	\kap = \left( \frac{\Theta}{8 \pi} \right)^{3/4} = \frac{3^{1/4}}{4 \pi \sqrt{\beta_{s}}} \per
\end{align}
We can also translate $\Delta \sigma_{\rm max}$ into an upper bound on the angle between ${\bf k}$ and ${\bf b}_{s}^{\prime}$:  
\begin{align}\label{eq:theta_max}
	\theta_{\rm max} = \frac{\Delta \sigma_{\rm max}}{L} = \Theta \, ( \kup L )^{-1/3}  \per
\end{align}

For the $\Ical_{-}$ integral, the analysis is similar, but the saddle point criterion is replaced with $k \cdot a_s^{\prime} = - \omega - {\bf k} \cdot {\bf a}_s^{\prime} = 0$ implying that ${\bf k} = - \kup \, {\bf a}_s^{\prime}$ at the quasi-saddle point.  
Consequently, in the equations analogous to \eref{eq:qsp_condit} all the signs on the right hand side are flipped.  
The results for both integrals can be summarized as 
\begin{align}\label{eq:I_sp}
	\Ical_{+} & \approx 
	A_{+} \frac{L \, b^{\prime}_{s}}{(\kup L)^{1/3}} 
	+ i B_{+} \frac{L^2 b^{\prime \prime}_{s}}{(\kup L)^{2/3}} 
	\Com
	\kap \, m \sqrt{m L} < \kup
	\Com 
	\theta_{k b_{s}^{\prime}} < \Theta \, (\kup L)^{-1/3}  \nn
	\Ical_{-} & \approx 
	A_{-} \frac{L \, a^{\prime}_{s}}{(\kup L)^{1/3}} 
	+ i B_{-} \frac{L^2 a^{\prime \prime}_{s}}{(\kup L)^{2/3}} 
	\Com
	\kap \, m \sqrt{m L} < \kup
	\Com 
	\theta_{k a_{s}^{\prime}} < \Theta \, (\kup L)^{-1/3} 
\end{align}
where $\theta_{k b_{s}^{\prime}}$ and $\theta_{k a_{s}^{\prime}}$ are the angles between ${\bf k}$ and ${\bf b}_s^{\prime}$ or ${\bf a}_{s}^{\prime}$, respectively.  
The dimensionless parameters are defined as 
\begin{align}\label{eq:AB_def}
	A_{\pm} = \frac{2\pi}{3\Gamma(2/3)} \left( \frac{3}{\pi^2 \gamma_{\pm}^2} \right)^{1/3} 
	\qquad {\rm and} \qquad
	B_{\pm} = \pm \frac{\Gamma(2/3)}{\sqrt{3}} \left( \frac{3}{\pi^2 \gamma_{\pm}^2} \right)^{2/3}  \Biggr|_{\shortstack{$\gamma_{+} = \beta_{s}$ \\ $\gamma_{-} = \alpha_s$}} \com
\end{align}
and the dimensionless shape parameters, $\beta_{s}$ and $\alpha_{s}$, are defined as in \eref{eq:shape_param}.  
For shorter wavelength radiation, $\kup < \kap \, m \sqrt{m L}$, there is no saddle point, and the integral vanishes rapidly.  
Also note that the approximations to $\Ical_{\pm}$ in \eref{eq:I_sp} satisfy the identities in \eref{eq:Ipm_iden} for ${\bf k} = \kup \, {\bf c}^{\prime}_{\pm}$ up to $O(m^2 / \kup^2)$ terms.

\subsection{Discontinuity Integral}\label{sub:Discontinuity}

In this appendix we will evaluate the integrals in \eref{eq:Ipm_def} for the case in which the gradient of the string worldsheet, $\partial_{\sigma} \Xbb^{\mu}$, has a discontinuity \cite{Damour:2001bk}.  
We first suppose that $b^{\mu}(\sigma_+)$ has $N_k$ discontinuities corresponding to $N_k$ kinks on the string loop.  
The typical distance between the kinks will be $D = L / N_k$.  
To calculate the contribution to $\Ical_{+}$ coming from a single discontinuity located at $\sigma_+ = \sigma_d$ we parametrize
\begin{align}\label{eq:b_discont}
	b^{\prime \, \mu}(\sigma_+)
	& =  
	\begin{cases}
	b_+^{\prime \, \mu} = \bigl\{ 1 \, , \, \hat{\bf m}_+ \bigr\}
	& \qquad 0 < \sigma_+ - \sigma_d < D/2 \\
	b_-^{\prime \, \mu} = \bigl\{ 1 \, , \, \hat{\bf m}_- \bigr\}
	& \qquad - D/2 < \sigma_+ - \sigma_d < 0 \\
	\end{cases}
\end{align}
where $\hat{\bf m}_{\pm}$ are unit vectors and $b_{\pm} = \bigl\{ \sigma_{+} \, , \, (\sigma_+ - \sigma_d) \, \hat{\bf m}_{\pm} \bigr\}$.  
Inserting \eref{eq:b_discont} into \eref{eq:Ipm_def} we approximate the worldsheet integral as 
\begin{align}\label{eq:Iplus_disc}
	\Ical_{+} 
	\approx \int_{-D/2}^{D/2} d\sigma_{+} \, b^{\prime} (\sigma_+) \, e^{i k \cdot b^{\prime} \sigma_+ / 2} 
	\approx & \ 
	\Biggl[ \frac{2}{\omega} \left ( \frac{b_{+}^{\prime}}{\hat k \cdot b_{+}^{\prime}} - \frac{b_{-}^{\prime}}{\hat k \cdot b_{-}^{\prime}} \right ) 
	\nn &
	- \frac{2}{\omega} \left ( \frac{b_{+}^{\prime}}{\hat k \cdot b_{+}^{\prime}} e^{i (\hat{k} \cdot b_{+}^{\prime}) \frac{\omega D}{4}}  - 
  \frac{b_{-}^{\prime}}{\hat k \cdot b_{-}^{\prime}} e^{-i (\hat{k} \cdot b_{-}^{\prime}) \frac{\omega D}{4}}  \right )  \Biggr]
	e^{i ( \frac{\omega \sigma_{d}}{2} + \pi ) }
\end{align}
where $\hat{k}^{\mu} \equiv k^{\mu} / \omega = \bigl\{ 1 \, , \, {\bf k} / \omega \bigr\}$.  
Upon integrating over the entire loop, the second term cancels among the contributions from different discontinuities (summing all kinks).  
Then we can drop both this second term and the overall phase to write the contribution from a single discontinuity as 
\begin{align}
	\Ical_{+} 
	\approx \frac{1}{\omega} \Bigl( \beta_{+} \, b_{+}^{\prime} - \beta_{-} \, b_{-}^{\prime} \Bigr) 
\end{align}
with $\beta_{\pm} \equiv 2 / (\hat{k} \cdot b_{\pm}^{\prime}) = 2 / (1 - \hat{\bf k} \cdot \hat{\bf m}_{\pm} )$.  

To calculate the integral $\Ical_{-}$ we parametrize $a^{\prime}(\sigma_-)$ in terms of $a^{\prime}_{\pm}$ in analogy with \eref{eq:b_discont}.  
We can summarize the results of both calculations as follows 
\begin{align}\label{eq:I_disc}
	\Ical_{+}(k) & \approx \frac{1}{\omega} \Bigl( \beta_{+} \, b_{+}^{\prime} - \beta_{-} \, b_{-}^{\prime} \Bigr)
	\ \Com \ \,
	\beta_{\pm} = \frac{2}{\hat{k} \cdot b_{\pm}^{\prime}}
	\ \Com \,
	L^{-1} < \omega
	\nn
	\Ical_{-}(k) & \approx \frac{1}{\omega} \Bigl( \alpha_{+} \, a_{+}^{\prime} - \alpha_{-} \, a_{-}^{\prime} \Bigr)
	\ \Com 
	\alpha_{\pm} = \frac{2}{\hat{k} \cdot a_{\pm}^{\prime}} 
	\ \Com 
	L^{-1} < \omega \per
\end{align}
The dimensionless coefficients are bounded as $1 \leq \beta_{\pm} , \alpha_{\pm}$.  
In the limit that $k$ coincides with one of the discontinuity vectors, $b_{\pm}^{\prime}$ or $a_{\pm}^{\prime}$, one finds that $\beta_{\pm}$ or $\alpha_{\pm} \to \infty$.  
This apparent divergence is an artifact of neglecting the second set of terms in \eref{eq:Iplus_disc}, and upon retaining these terms one can see that $\Ical_{+} \sim D \ll 1 / \omega$ in the limit that $(\hat{k} \cdot b_{+}^{\prime}) \omega D \ll 1$.  
Therefore we must restrict ourselves to the regime $\omega > D^{-1} \sim N_k L^{-1}$ and where $k \cdot b_{\pm}^{\prime}$ is away from zero; it follows that $1 \leq \beta_{\pm} , \alpha_{\pm} \lesssim {\rm few}$.  
To properly treat the case $k \cdot b_{+}^{\prime} = 0$ in which the phase is stationary, one should use the saddle point approximation, as described in \sref{sub:SaddlePoint}.

\section{Scalar and Tensor Integrals for Cusps, Kinks, and Kink Collisions}\label{app:ST_integrals}

Here we evaluate the scalar and tensor integrals, $\Ical$ and $\Ical^{\mu \nu}$ given by \erefs{eq:IcalH}{eq:IcalZ}, for the cusp, kink, and kink-kink collision string configurations.  
For the scalar integral, we will only be interested in the modulus $| \Ical |^2$.  
For the tensor integral, we will only be interested in the (positive, real) scalar combinations
\begin{align}
	\Pi(q) & = - g_{\nu \beta} \, \frac{q_{\mu} q_{\alpha}}{q^2} \, \Ical^{\mu \nu}(q) \Ical^{\alpha \beta}(q)^{\ast} 
	\\
	\Upsilon(k, \bar{k}) & = \frac{2 k_{\mu} \bar{k}_{\nu}}{(k+\bar{k})^2} \, \Ical^{\mu \nu}( k + \bar{k} )
\end{align}
which were originally defined in \erefs{eq:Pi_def}{eq:Upsilon_def}.  

We can simply the expression for $\Pi(q)$ as follows.  
Using \eref{eq:IcalZ} and the identity 
\begin{align}
	(-g_{\nu \beta}) \epsilon^{\mu \nu \gamma \delta} \epsilon^{\alpha \beta \rho \sigma} 
	= & \   
	g^{\mu \alpha} g^{\gamma \rho} g^{\delta \sigma} 
	+ g^{\mu \rho} g^{\gamma \sigma} g^{\delta \alpha} 
	+ g^{\mu \sigma} g^{\gamma \alpha} g^{\delta \rho} \nn 
	& 
	- g^{\mu \alpha} g^{\gamma \sigma} g^{\delta \rho} 
	- g^{\mu \rho} g^{\gamma \alpha} g^{\delta \sigma} 
	- g^{\mu \sigma} g^{\gamma \rho} g^{\delta \alpha} 
\end{align}
we can write $\Pi$ as 
\begin{align}
	\Pi(q) & = 
	\frac{1}{4} \Bigl[ ( \Ical_{+} \cdot \Ical_{+}^{\ast} ) ( \Ical_{-} \cdot \Ical_{-}^{\ast} ) - | \Ical_{+} \cdot \Ical_{-}^{\ast} |^2 \Bigr]
	+ \frac{1}{2} \frac{1}{q^2} {\rm Re} \, \Bigl[ (q \cdot \Ical_{+}^{\ast}) (q \cdot \Ical_{-}) (\Ical_{+} \cdot \Ical_{-}^{\ast} ) \Bigr] 
	\nn & \qquad
	- \frac{1}{4} \frac{1}{q^2} \Bigl[ (q \cdot \Ical_{+})(q \cdot \Ical_{+}^{\ast}) ( \Ical_{-} \cdot \Ical_{-}^{\ast} ) + ( q\cdot \Ical_{-} ) ( q\cdot \Ical_{-}^{\ast} ) ( \Ical_{+} \cdot \Ical_{+}^{\ast} )  \Bigr] \per
\end{align}
Furthermore, from the periodicity of the string worldsheet, we have the identity $q \cdot \Ical_{\pm}(q) = 0$ [see \eref{eq:Ipm_iden}].  
Making this simplification we finally obtain 
\begin{align}
	\label{eq:Pi}
	\Pi(q) = 
	\frac{1}{4} \Bigl[ ( \Ical_{+} \cdot \Ical_{+}^{\ast} ) ( \Ical_{-} \cdot \Ical_{-}^{\ast} ) - | \Ical_{+} \cdot \Ical_{-}^{\ast} |^2 \Bigr] 
	\per 
\end{align}

We can simplify the expression for $\Upsilon(k,\bar{k})$ as follows.  
Let $q = k + \bar{k}$ and $p = k - \bar{k}$.  
Using \eref{eq:IcalZ} we can express $\Upsilon$ as 
\begin{align}\label{eq:Upsilon}
	2 \Upsilon(k,\bar{k}) \, q^2 
	= \ & p_{\mu} q_{\nu} \Ical_{+ \alpha}(q) \Ical_{- \beta}(q) \epsilon^{\mu \nu \alpha \beta}
	\nn
	= \ &  - p^{0} {\bf q} \cdot ( {\bm \Ical}_{+} \times {\bm \Ical}_{-} ) 
	+ q^{0} {\bf p} \cdot ( {\bm \Ical}_{+} \times {\bm \Ical}_{-} )  \nn
	& - \Ical_{+}^{0} \, {\bf p} \cdot ( {\bf q} \times {\bm \Ical}_{-} ) 
	+ \Ical_{-}^{0} \, {\bf p} \cdot ( {\bf q} \times {\bm \Ical}_{+})
	\per
\end{align}
Using the identities $q \cdot p = q \cdot \Ical_{+} = q \cdot \Ical_{-} = 0$, this can also be written as \cite{Chu:2010zzb}
\begin{align}\label{eq:Upsilon_v2}
	\Upsilon(k, \bar{k}) = \frac{1}{2 q^0} \, {\bf p} \cdot ( {\bm \Ical}_{+} \times {\bm \Ical}_{-} )  \per 
\end{align}

To compare $\Upsilon$ with $\Pi$, it is convenient to move to the frame in which $q^{\mu} = \bigl\{ q^0 \, , \, {\bf 0} \bigr\}$.  
Then the identities $q \cdot p = q \cdot \Ical_{+} = q \cdot \Ical_{-} = 0$ require $p$, $\Ical_{+}$, and $\Ical_{-}$ to have vanishing time-like components.  
In this frame, we can write 
\begin{align}
	\Pi(q) = \frac{ | {\bm \Ical}_{+} | \, | {\bm \Ical}_{-} | }{4} \sin^2 (\theta_{+-})
\end{align}
where $\theta_{+-}$ is the angle between ${\bm \Ical}_{+}$ and ${\bm \Ical}_{-}$.  
Further denoting $\theta_{p+-}$ as the angle between ${\bf p}$ and ${\bm \Ical}_{+} \times {\bm \Ical}_{-}$ we have
\begin{align}
	|\Upsilon(k, \bar{k})|^2 =  \left[ \left( \frac{| {\bf p} |}{q^0} \right)^2 \cos^2 (\theta_{p+-}) \right] \frac{ | {\bm \Ical}_{+} |^2 \, | {\bm \Ical}_{-} |^2 }{4} \sin^2 (\theta_{+-}) \per
\end{align}
The two expressions are related by 
\begin{align}\label{eq:Upsilon_to_Pi}
	|\Upsilon(k, \bar{k})|^2 = \left[ \left( \frac{| {\bf k} - \bar{\bf k} |}{\omega + \bar{\omega}} \right)^2 \cos^2 (\theta_{p+-}) \right] \Pi(q) 
\end{align}
where the quantity is square brackets is always $\leq 1$.  
The inequality is saturated when ${\bf p}$ is aligned with ${\bm \Ical}_{+} \times {\bm \Ical}_{-}$ (\ie, $\theta_{p+-} \approx 0$) and either $\kup \gg \bar{\kup}$ or $\bar{\kup} \gg \kup$.

\subsection{Scalar Integral -- Cusp}\label{sub:ScalarCusp}

A cusp occurs when both integrals $\Ical_{+}(b;q)$ and $\Ical_{-}(a;q)$ have a saddle point at the same value of $q^{\mu}$ [see \eref{eq:qsp_condit}].  
This requires ${\bf q} = \qup \, {\bf b}^{\prime}_{c} = - \qup \, {\bf a}^{\prime}_{c}$ or equivalently 
\begin{align}\label{eq:cusp_crit}
	a_{\rm c}^{\prime} = - b^{\prime}_{c} \per
\end{align}
The scalar integral, $\Ical$ from \eref{eq:IcalH}, is evaluated using the expressions for $\Ical_{\pm}$ in \eref{eq:I_sp}.  
Using the identities from \eref{eq:12prime_iden} most of the four-vector contractions vanish.  
The surviving term is proportional to ${\bf a}_{c}^{\prime \prime} \cdot {\bf b}_{c}^{\prime \prime} = (2\pi/L)^2 \alpha_{c} \beta_{c} \, \hat{\bf a}_{c}^{\prime \prime} \cdot \hat{\bf b}_{c}^{\prime \prime}$ where we have used shape shape parameters, introduced in \eref{eq:shape_param}.  
Then, the squared integral evaluates to
\begin{align}\label{eq:I_scalar_cusp}
	\bigl| \Ical^{\rm (cusp)}(q) \bigr|^2
	& = 
	\Scusp 
	\frac{L^{4/3}}{\qup^{8/3}} 
	\Com 
	\kap \, m \sqrt{mL} < \qup
	\Com 
	\theta < \Theta \, (\qup L)^{-1/3} \, {\rm (cone)}
\end{align}
where $\Scusp \equiv \pi^4 \alpha_c^2 \beta_c^2 B_{-}^2 B_{+}^2 \cos^2 \theta_{ab}$ with $B_\pm$ defined in \eref{eq:AB_def}, and where $\theta_{ab}$ is the angle between ${\bf a}_{c}^{\prime \prime}$ and ${\bf b}_{c}^{\prime \prime}$.  
The angle between ${\bf k}$ and ${\bf b}_{c}^{\prime} = - {\bf a}_{c}^{\prime}$ is bounded above by the saddle point criterion, and therefore ${\bf k}$ falls within a cone of opening angle $\Theta (\qup L)^{-1/3}$ centered on the cusp.  

The dimensionless prefactor, $\Scusp$, may be estimated using the expressions for $B_{\pm}$ in \eref{eq:AB_def}.  
The shape parameters, $\alpha_c$ and $\beta_c$, are expected to be $O(1)$, but their precise values cannot be determined without greater knowledge of the nature of the cusp.  
In order to track how this uncertainty in the magnitude of the shape parameter feeds into the particle production calculation, we will consider a fiducial range of values for $\alpha_c$ and $\beta_c$.  
Estimating $1/5 \lesssim \alpha_c , \beta_c \lesssim 5$ and $\cos \theta_{ab} \approx 1$ we find $0.2 \lesssim \Scusp \lesssim 10$.  
The dimensionless parameters $\kap$ and $\Theta$, given by \erefs{eq:kmin_bound}{eq:theta_max}, are less sensitive to the uncertainty in the shape parameters.  
Typically $\kap \approx 0.01$ and $\Theta \approx 0.1$.  

\subsection{Scalar Integral -- Kink}\label{sub:ScalarKink}

A kink occurs when the derivative of one of the functions $b^{\mu}(\sigma_+)$ or $a^{\mu}(\sigma_-)$ appearing in $\Ical_{+}(b;q)$ or $\Ical_{-}(a;q)$ has a discontinuity, and the other integral has a saddle point.  
For the sake of discussion we suppose that $\Ical_+$ contains the saddle point and $\Ical_-$ the discontinuity. 
We calculate $\Ical$ by inserting \erefs{eq:I_sp}{eq:I_disc} in \eref{eq:IcalH}.  
From \eref{eq:I_sp} we see that the leading order term in $\Ical_{+}$ is proportional to $b_s^{\prime}$ and the subleading term is proportional to $b_{s}^{\prime \prime}$.  
Upon contracting with $\Ical_{-}$ the leading order term is negligible:  we have the identity $q \cdot \Ical_{-} =0$ [\eref{eq:Ipm_iden}] and the saddle point criterion ${\bf q} = \qup {\bf b}_{s}^{\prime}$ from which it follows that $b_{s}^{\prime} \cdot \Ical_{-} = - (q^0 / \qup - 1) \Ical_{-}^{0} \approx - (m^2 / 2 \qup^2) \Ical_{-}^{0}$, which is negligible (at $\qup > m \sqrt{mL}$) compared to the terms that we keep.\footnote{
The result of \rref{Lunardini:2012ct} is derived using this ``leading order'' term, $\Ical_{+} \sim b_{s}^{\prime}$.  
}  

The calculation described above yields 
\begin{align}\label{eq:Ih_kink}
	 \Ical^{\rm (kink)}(q) =  
	-i \frac{B_{+}}{4} \frac{L^{4/3}}{\qup^{5/3}} \bigl[ \alpha_{+} \, ( b_{s}^{\prime \prime} \cdot a_{+}^{\prime}) - \alpha_{-} \, ( b_{s}^{\prime \prime} \cdot a_{-}^{\prime}) \bigr]
\end{align}
Here we have used $q^0 \approx \qup$ since the saddle point condition requires $m \sqrt{mL} < \qup$ and $m L \gg 1$ for typical size loops.  
For the same reason, the bound on the discontinuity integral, $L^{-1} < \qup$, is subsumed by the bound on the saddle point integral, $m \sqrt{mL} < \qup$.  
The squared integral becomes
\begin{align}
\label{eq:I_scalar_kink}
	\bigl|  \Ical^{\rm (kink)}(q) \bigr|^2 =  
	\Skink \frac{L^{2/3}}{\qup^{10/3}}
	\Com 
	\kap \, m \sqrt{mL} < \qup
	\Com
	\theta < \Theta \, (\qup L)^{-1/3} \, {\rm (band)}
\end{align}
where 
$\Skink \equiv \frac{(2\pi)^2}{16} B_{+}^2 \beta_{s}^2 \bigl[ \alpha_{+} \, ( \hat{\bf b}_{s}^{\prime \prime} \cdot {\bf a}_{+}^{\prime}) - \alpha_{-} \, ( \hat{\bf b}_{s}^{\prime \prime} \cdot {\bf a}_{-}^{\prime}) \bigr]^2$
and we have used the shape parameters, introduced in \eref{eq:shape_param}.  
The saddle point criterion requires ${\bf q}$ to be aligned with ${\bf b}_{s}^{\prime}$.  
Consequently, the radiation is emitted into a band (whose orientation is determined by ${\bf b}_{s}^{\prime}(\sigma_+)$) of angular width $\Theta (\qup L)^{-1/3}$ and angular length $\sim 2\pi$.  

We can estimate a range of uncertainty for $\Skink$ as we did in \aref{sub:ScalarCusp}.  
Recall that $B_+$ was given by \eref{eq:AB_def}.  
Following the convention established in \aref{sub:ScalarCusp}, we estimate the shape parameter as $1/5 \lesssim \beta_s \lesssim 5$.  
We also take $1 \lesssim \alpha_{\pm} \lesssim 5$, as per the discussion below \eref{eq:I_disc}.  
Together this lets us estimate $0.1 \lesssim \Skink \lesssim 20$

\subsection{Scalar Integral -- Kink Collision}\label{sub:ScalarKinkKink}

For the case of a kink-kink collision both integrals, $\Ical_+$ and $\Ical_-$, have discontinuities and are given by \eref{eq:I_disc}.  
The scalar integral is evaluated from \eref{eq:IcalH} to be 
\begin{align}\label{eq:Ih_kinkkink}
	\Ical^{\rm (k-k)}(q) = - \frac{1}{4\omega^2} \Bigl[ 
	& (b_{+}^{\prime} \cdot a_{+}^{\prime}) (\beta_{+} \alpha_{+}) 
	- (b_{+}^{\prime} \cdot a_{-}^{\prime}) (\beta_{+} \alpha_{-}) \nn
	& \ - (b_{-}^{\prime} \cdot a_{+}^{\prime}) (\beta_{-} \alpha_{+}) 
	+ (b_{-}^{\prime} \cdot a_{-}^{\prime}) (\beta_{-} \alpha_{-}) 
	\Bigr] 
\end{align}
where $\omega = q^0$.  
The square is 
\begin{align}\label{eq:I_scalar_kk}
	\bigl|  \Ical^{\rm (k-k)} \bigr|^2 & = \frac{\Skk}{\omega^4} 
	\Com
	L^{-1} < \omega 
	\per
\end{align}
We have defined $\Skk \equiv \frac{1}{16} \, \left[ \sum \pm ( 1 + {\bf b}_{\pm}^{\prime} \cdot {\bf a}_{\pm}^{\prime} ) \beta_{\pm} \alpha_{\pm} \right]^2$ where the sum runs over all possible combinations of $+$ and $-$ as given by \eref{eq:Ih_kinkkink}.  
For the case of a discontinuity, the worldsheet integrals, $\Ical_{\pm}$, are insensitive to the orientation of ${\bf k}$ (see \sref{sub:Discontinuity}) and the corresponding radiation is emitted approximately isotropically.  

Recall that $\alpha_{\pm}$ and $\beta_{\pm}$ were given by \eref{eq:I_disc}, and following the conventions established in \aref{sub:ScalarKink}, we estimate $1 \lesssim \beta_{\pm} , \alpha_{\pm} \lesssim 5$.  
This yields the estimate $1 \lesssim \Skk \lesssim 500$.

\subsection{Tensor Integral -- Cusp}\label{sub:TensorCusp}

If both $\Ical_{\pm}$ contain a saddle point, then we evaluate the tensor integral by inserting \eref{eq:I_sp} into \eref{eq:Pi}.  
After making use of the identities in \eref{eq:12prime_iden}, many of the terms vanish leaving only 
\begin{align}
	\Pi(q) = \frac{B_{+}^2 B_{-}^2 L^8}{4 (\qup L)^{8/3}} \Bigl[ (b_{c}^{\prime \prime} \cdot b_{c}^{\prime \prime}) (a_{c}^{\prime \prime} \cdot a_{c}^{\prime \prime}) - (a_{c}^{\prime \prime} \cdot b_{c}^{\prime \prime})^2 \Bigr] \per
\end{align}
We extract the factors of $L$ from the bracketed quantities by using the parametrization in \eref{eq:shape_param}.  
Doing so gives 
\begin{align}\label{eq:Pi_cusp}
	\Pi(q) \bigr|^{\rm (cusp)}
	= \Tcusp \frac{L^{4/3}}{\qup^{8/3}}
	\Com
	\kap \, m \sqrt{mL} < \qup 
	\Com
	\theta < \Theta \, (\qup L)^{-1/3} \, {\rm (cone)}
\end{align}
where $\Tcusp \equiv \frac{(2\pi)^4}{4} (B_{+} B_{-})^2 \alpha_c^2 \beta_c^2 \sin^2 \theta_{a b}$ and $\theta_{ab}$ is the angle between ${\bf a}_{c}^{\prime \prime}$ and ${\bf b}_{c}^{\prime \prime}$.  
The angle between ${\bf q}$ and ${\bf b}_{c}^{\prime} = - {\bf a}_{c}^{\prime}$ is bounded above by $\Theta (\qup L)^{-1/3}$, and consequently ${\bf q}$ is oriented within a cone centered at the cusp.  

We can estimate the dimensionless coefficient by making the same estimates as in \aref{sub:ScalarCusp}.  
Assuming that the shape parameters fall into the range $1 / 5 < \alpha_c , \beta_c < 5$ and approximating $(1-\cos^2 \theta_{ab}) \approx 1$ we obtain $0.5 \lesssim \Tcusp \lesssim 50$.  

\subsection{Tensor Integral -- Kink}\label{sub:TensorKink}

If $\Ical_{+}$ contains a saddle point and $\Ical_{-}$ contains a discontinuity, then we evaluate the tensor integral by inserting \erefs{eq:I_sp}{eq:I_disc} into \eref{eq:Pi}.  
Some of the contractions vanish upon using the identities in \eref{eq:12prime_iden}.  
As we discussed in \sref{sub:ScalarKink}, the leading order term in $\Ical_{+}$ is negligible because the contraction $b_{s}^{\prime} \cdot \Ical_{-}$ is suppressed by $m^2 / \qup^2 \ll 1$.  
Making these substitutions we are left with 
\begin{align}
	\Pi(q) = \frac{1}{4} \Biggl[
	\frac{B_{+}^2 L^4 (b_{s}^{\prime \prime} \cdot b_{s}^{\prime \prime})}{(\qup L)^{4/3}} \frac{(-2) \, \alpha_{+} \alpha_{-} (a_{+}^{\prime} \cdot a_{-}^{\prime})}{\qup^2} 
	- \frac{1}{\qup^2} \frac{B_{+}^2 L^2 ( b_{s}^{\prime \prime} \cdot A^{\prime} )^2 }{(\qup L)^{4/3}}
	\Biggr]
\end{align}
where $A^{\prime} \equiv \alpha_{+} a_{+}^{\prime} - \alpha_{-} a_{-}^{\prime}$.  
We relate $b_{s}^{\prime \prime}$ to $\beta_s$ using the parametrization in \eref{eq:shape_param}.  
Then 
\begin{align}\label{eq:Pi_kink}
	\Pi(q) \bigr|^{\rm (kink)} = 
	\Tkink \frac{L^{2/3}}{\qup^{10/3}}
	\Com
	\kap \, m \sqrt{mL} < \qup 
	\Com
	\theta < \Theta \, (\qup L)^{-1/3} \, {\rm (band)}
\end{align}
where 
\begin{align}
	\Tkink \equiv \frac{(2\pi)^2}{2} B_{+}^2 \beta_{s}^2
	\Biggl( &
	\alpha_{+} \alpha_{-}
	\Bigl( 
	 (1 - {\bf a}_{+}^{\prime} \cdot {\bf a}_{-}^{\prime}) 
	 +
	 (\hat{\bf b}_{s}^{\prime \prime} \cdot {\bf a}_{+}^{\prime}) (\hat{\bf b}_{s}^{\prime \prime} \cdot {\bf a}_{-}^{\prime})
	 \Bigr) 
	 \nn & 
	- \frac{ \alpha_{+}^2 }{2} (\hat{\bf b}_{s}^{\prime \prime} \cdot {\bf a}_{+}^{\prime})^2
	- \frac{ \alpha_{-}^2 }{2} (\hat{\bf b}_{s}^{\prime \prime} \cdot {\bf a}_{-}^{\prime})^2
	\Biggr) \per
\end{align}
The momentum ${\bf k}$ is constrained to fall within a band of angular width $\Theta (\kup L)^{-1/3}$.  

Following the conventions from the previous sections, we estimate $1/5 \lesssim \beta_s \lesssim 5$ and determine $B_+$ from \eref{eq:AB_def}.  
We estimate the parenthetical factor as simply $|\alpha_+ \alpha_-|$ and take $1 \lesssim \alpha_{\pm} \lesssim 5$ as before.  
Then together we find $1 \lesssim \Tkink \lesssim 200$.

\subsection{Tensor Integral -- Kink Collision}\label{sub:TensorKinkKink}

If both $\Ical_{\pm}$ possess a discontinuity point, then we evaluate the tensor integral by inserting \eref{eq:I_disc} into \eref{eq:Pi}.  
This gives 
\begin{align}
	\Pi(q) & = \frac{1}{4(q^0)^4} \Bigl[ ( B^{\prime} \cdot B^{\prime} ) ( A^{\prime} \cdot A^{\prime}) - ( B^{\prime} \cdot A^{\prime} )^2 \Bigr]
\end{align}
where $B^{\prime} \equiv \beta_{+} b_{+}^{\prime} - \beta_{-} b_{-}^{\prime}$ and $A^{\prime} \equiv \alpha_{+} a_{+}^{\prime} - \alpha_{-} a_{-}^{\prime}$.  
This can be written as 
\begin{align}\label{eq:Pi_kk}
	\Pi(k) \bigr|^{\rm (k-k)}
	= \Tkk \frac{1}{(q^0)^4}
	 \Com
	L^{-1} < q^0
\end{align}
where $\Tkk \equiv \bigl[ \beta_{+} \beta_{-} \alpha_{+} \alpha_{-} \, (b_{+}^{\prime} \cdot b_{-}^{\prime})  (a_{+}^{\prime} \cdot a_{-}^{\prime}) - \frac{1}{4} [ (\beta_{+} b_{+}^{\prime} - \beta_{-} b_{-}^{\prime}) \cdot (\alpha_{+} a_{+}^{\prime} - \alpha_{-} a_{-}^{\prime}) ]^2 \bigr]$, and we have used that $a_{\pm}^{\prime}$ and $b_{\pm}^{\prime}$ are null vectors.  

We can estimate $\Tkk$ following the conventions established in \aref{sub:ScalarKinkKink}.  
We take $1 \lesssim \alpha_{\pm} , \beta_{\pm} \lesssim 5$ and approximate $b_{+}^{\prime} \cdot b_{-}^{\prime} \approx a_{+}^{\prime} \cdot a_{-}^{\prime} \approx b_{\pm}^{\prime} \cdot a_{\pm}^{\prime} \approx 1$.  
This allows us to estimate the range $0.2 \lesssim \Tkk \lesssim 200$ for the dimensionless coefficient.

\section{Cusp Boost Factor and UV Sensitivity}\label{app:Boost}

It was recognized in \rref{Vachaspati:2009kq} that particle radiation from a cusp will be highly boosted  since the cusp tip moves at the speed of light in the rest frame of the loop.  
(By contrast, the gravitational radiation spectrum is IR sensitive, and the boost factor is not relevant.)  
At a given point on the string loop, the boost factor is given by 
\begin{align}\label{eq:boost_def}
	\gamma_{\rm boost}(\tau,\sigma)
	= \frac{1}{ \sqrt{ \dot{\Xbb}^{\mu}(\tau,\sigma) \dot{\Xbb}_{\mu}(\tau,\sigma) } } 
	= \sqrt{ \frac{-2}{ a^{\prime}(\sigma - \tau) \cdot b^{\prime}(\sigma + \tau) } }
\end{align}
where we have used the formulae in \aref{app:Worldsheet}.  
Expanding both $a^{\mu}$ and $b^{\mu}$ as in \eref{eq:phase} and using \eref{eq:12prime_iden} gives 
\begin{align}
	\gamma_{\rm boost}(\Delta \sigma) = \frac{(L / \Delta \sigma)}{\pi \sqrt{\alpha_s^2 + \beta_s^2}}
\end{align}
where $\Delta \sigma$ is the distance from the tip of the cusp.  
The dimensionless shape parameters, $\beta_s$ and $\alpha_s$, were defined in \eref{eq:shape_param}.  
The boost factor grows with decreasing $\Delta \sigma$ as one investigates radiation coming from closer and closer to the tip of the cusp.  
For a ideal string of zero thickness, we can take $\Delta \sigma \to 0$ and $\gamma_{\rm boost} \to \infty$.  
In reality, the finite thickness string overlaps with itself at the cusp tip, and a segment of string with length $\Delta \sigma_{\rm min} \sim \sqrt{L / M}$ will evaporate into particle radiation \cite{Olum:1998ag}.  
This leads to an upper bound on the boost factor, 
\begin{align}\label{eq:boost}
	\gamma_{\rm boost} \lesssim \frac{\sqrt{ML}}{\pi \sqrt{\alpha_s^2 + \beta_s^2}} \per
\end{align}
The radiation spectra that we calculate should drop off when the momentum of the radiated particle exceeds the inverse string thickness.  
In the rest frame of the radiating string segment this condition is $\kup_{\rm cusp-frame} < M$, but in the rest frame of the loop this condition is $\kup_{\rm loop-frame} < M \gamma_{\rm boost}$.  
For radiation from a cusp this becomes $\kup_{\rm loop-frame} < M \sqrt{ML}$ whereas for radiation from a (non-relativistic) kink this becomes $\kup_{\rm loop-frame} < M$.  

Since $\kup$ exceeds $M$ for radiation from a cusp, one may worry that the effective field theory assumption has broken down.  
Upon ``integrating out'' the heavy string degrees of freedom, $S$ and $X^{\mu}$, we dropped the infinite tower of higher-order, non-renormalizable interactions between the light SM fields and the string worldsheet.  
For example, in \eref{eq:Psi_Leff} we consider the non-renormalizable interaction of SM fermions with the string worldsheet, but we do not treat higher-order operators such as $\frac{C}{M^4} \, [ \bar{\Psi}(x) \Psi(x) ]^2 \, \int d^2 \sigma \, \sqrt{-\gamma} \, \delta^{(4)}(x - \Xbb)$.  
These operators will also contribute to the radiation spectrum, but their contribution will be proportional to some power of the ratio $k^2 / M^2 = \bar{k}^2 / M^2 = m^2 / M^2$ or $(k \cdot \bar{k}) / M^2$.  
Then as long as these ratios are small compared to one, we can neglect the higher order operators.  
Throughout the paper, we have assumed that $m_H \sim m_Z \sim m_{\psi} \ll M^2$, and the first condition is satisfied.  
The second ratio can be written as $k \cdot \bar{k} / M^2 \approx (\kup \bar{\kup} / M^2) ( 1 - \cos \theta_{k \bar{k}} )$ where $\theta_{k \bar{k}}$ is the angle between ${\bf k}$ and $\bar{\bf k}$.  
Assuming $\kup \approx \bar{\kup}$ and $\theta_{k\bar{k}} \ll 1$ this bound becomes $\kup \theta_{k \bar{k}} \ll M$.  
For isotropic radiation, as in the case of a kink, we need $\kup \ll M$.  
On the other hand, if the radiation is beamed into a cone, then we can have $\kup \gg M$ (in the rest frame of the loop) without invalidating the effective field theory analysis.  
Specifically, for radiation from a cusp we have $\theta_{k \bar{k}} < \Theta \, (\kup L)^{-1/3}$ and $\kup < M \sqrt{ML}$, which together satisfy $\kup \theta_{k \bar{k}} < M$.

\end{appendix}


\end{document}